\documentclass[11pt]{article}
\pdfoutput=1

\usepackage{verbatim}
\usepackage{amsmath}     

\usepackage{amssymb}					
\DeclareSymbolFontAlphabet{\amsmathbb}{AMSb}

\usepackage{simplewick} 
\usepackage{mathtools}

\usepackage{enumerate}
\usepackage{enumitem}

\usepackage{braket}
\usepackage{cancel}

\usepackage{latexsym}
\usepackage{lscape} 
\usepackage{epsfig}
\usepackage{color}
\usepackage{graphicx}
\usepackage[table]{xcolor}
\usepackage{enumerate}
\usepackage{hhline}
\usepackage{subfig}
\usepackage{multirow}
\usepackage{mathrsfs}
\usepackage{dsfont}
\usepackage{longtable}

\usepackage{amscd}
\usepackage{cite}

\usepackage[colorlinks]{hyperref}
\hypersetup{%
  colorlinks = false,
  linkcolor  = red
}
\usepackage[
color=yellow]{todonotes}

\usepackage{pdfpages}

\usepackage{xspace}

\DeclareMathSymbol{\mg}{\mathrel}{symbols}{"1D}

\newcommand{\ga}{\alpha}
\newcommand{\gb}{\beta}

\newcommand{\gl}{\lambda}

\newcommand{\gs}{\sigma}

\newcommand{\beq}{\begin{equation}}
\newcommand{\eeq}{\end{equation}}
\newcommand{\barr}{\begin{array}}
\newcommand{\earr}{\end{array}}
\newcommand{\equ}[1]{\begin{gather} #1 \end{gather}}

\newcommand{\sfrac}[2]{\mbox{$\frac{#1}{#2}$}}
\newcounter{oldcounter}

\newcommand{\qand}{\quad \text{and} \quad}

\newcommand{\ba}[2]{\[\begin{array}{#2}\label{#1}}
\newcommand{\ea}{\end{array}\]}
\newcommand{\be}{\begin{equation}}
\newcommand{\ee}{\end{equation}}
\newcommand{\bea}{\begin{eqnarray}}
\newcommand{\eea}{\end{eqnarray}}

\usepackage{tikz}
\usetikzlibrary{calc}

\newcommand{\e}{{\text{e}}}

\newcommand{\order}[1]{\mathcal O\left(#1\right)}

\newcommand{\abs}[1]{\left\vert#1\right\vert} 

\newcommand{\quads}[1]{\quad #1 \quad}

\usepackage{tikz}

\newcommand{\vev}{\textsc{vev}\xspace}

\newcommand{\VEV}[1]{\langle#1\rangle}

\newcommand{\round}[1]{\left\lfloor#1\right\rceil}

\newcommand{\vset}{\textbf v}
\newcommand{\hset}{\textbf h}
\newcommand{\hzset}{\textbf h^z}

\newcommand{\rbm}{\textsc{rbm} }
\newcommand{\vrbm}{v\textsc{rbm} }

\newcommand{\weight}{\omega}
\newcommand{\bias}{\beta}
\newcommand{\obias}{\xi} 
\newcommand{\feq}{\mathcal E} 

\usepackage{appendix}
\usepackage{chngcntr}
\usepackage{etoolbox}
\usepackage{lipsum}
\AtBeginEnvironment{subappendices}{%
\chapter*{Appendix}
\addcontentsline{toc}{chapter}{Appendices}
\counterwithin{figure}{section}
\counterwithin{table}{section}
}

\begin{document}


\newcommand{\CreateTitlePage}
{
\thispagestyle{empty}

\begin{flushright}
\phantom{SBA-xxx}
\end{flushright}
\vskip 2 cm
{\Large {\bf 
Self-regularizing \\[1ex]
restricted Boltzmann machines
} 
}
\\[0pt]

\bigskip
\bigskip {\large
{\bf 
{Orestis Loukas}}\footnote{
E-mail: OLoukas@sba-research.org}
\bigskip }\\[0pt]
\vspace{0.23cm}
{\it 
SBA Reasearch   \\[0.5ex] 
Floragasse 7, 1040 Wien, Austria
}
%
\\[1ex] 
\bigskip
}

%

\thispagestyle{empty}
\begin{center}
\CreateTitlePage
\end{center}

\vspace{3cm}
\subsection*{\centering Abstract}

Focusing on the grand-canonical extension of the ordinary restricted Boltzmann machine, we suggest an energy-based model for feature extraction that uses a layer of hidden units with varying size.
By an appropriate choice of the chemical potential and given a sufficiently large number of hidden resources the generative model  is able to efficiently deduce the optimal number of hidden units required to learn the target data with exceedingly small generalization error. 
The formal simplicity  of the grand-canonical ensemble combined with a rapidly converging ansatz in mean-field theory enable us to recycle well-established numerical algothhtims during training, like contrastive divergence, with only minor changes.
As a proof of principle and to demonstrate the novel features of grand-canonical Boltzmann machines, we train our generative models on data from the Ising theory and  \textsc{mnist}.

\newpage

\tableofcontents


\section{Introduction}

In the past decades, artificial intelligence has increasingly become a major key-player in a vastly wide range of fields.
Training a machine to recognize patterns through versatile data, perform classification tasks and make decisions has been proven most of the times particularly successful, quite often outperforming hard-coded programs and  human cognition. 
Also within physics, the implementation of machine learning (\textsc{ml})  proves beneficial.
%
The related literature ranges from (un)supervised leaning on statistical  systems (for a concise introductory review see~\cite{2018arXiv180308823M}), many-body problems~\cite{Carleo602,2019arXiv190506034V} and quantum entanglement~\cite{2017arXiv170401552L,PhysRevX.7.021021} up to high-energy applications in Particle Phenomenology (e.g.~\cite{komiske2017deep,roe2005boosted,baldi2014searching}), String Theory (e.g.~\cite{ruehle2017evolving,he2017deep,he2018calabi}) and holography~\cite{hashimoto2018deep,gan2017holography,you2018machine,inproceedings}.

Yet, there are situations where the machine either after seemingly appropriate  training unexpectedly  fails to produce an adequate output or, to begin with, cannot learn the given data, at all.
The often unpredicted failure of the intelligent algorithms as well as their surprising success at specific tasks signify our lack of a concrete understanding of the theory underlying most of \textsc{ml} applications.
At this point, input from theoretical physics can be proven beneficial. 
%
Among the various ideas invoked in the interface between the theoretical description of physical systems and machine learning  to interpret and improve (deep) learning algorithms geometrization~\cite{2019arXiv190102354D}, variational approaches~\cite{unknown,2014arXiv1410.3831M,2016PhRvB..94p5134T} and classical thermodynamics~\cite{2017NJPh...19k3001G,Tubiana:2016zpw} have been proposed.
In \cite{2017arXiv170804622M,Cossu:2018pxj,Iso:2018yqu,Funai:2018esm,2018arXiv180909632L,2018NatPh..14..578K} ideas from the Renormalization Group flow are used to comprehend the flow of configurations triggered by generative models after training on systems from condensed matter physics.

Besides the concrete model and  type of task performed (classification vs.\ generative), a failure of the \textsc{ml} algorithm to recognize the desired features from the target data is intimately related to the existence of various so called \textit{hyperparameters} which  require  fine-tuning before or even during training. 
Most of those hyperparameters concern the architecture of the \textsc{ml} model. This comprises the depth of the neural network,  the number of units at each hidden layer and the activation function(s) used.
In contrast to  hyperparameters, other parameters of the model like the weights and biases are determined during training by extremizing an appropriate information-theoretic metric~\cite{1957PhRv..106..620J,renyi1961measures} like cross entropy which conventionally measures how well the model can classify or reproduce the given data.
 
Generally, deeper networks tend to extract features from a target system with a higher level of sophistication.
Similarly, hidden layers  of bigger size can approximate functions with increasing accuracy and thus help to learn better a provided data set. 
However, opting for larger architectures comes at a price. Besides computational efficiency deeper networks sometimes bring no advantage over shallow models~\cite{2017arXiv170804622M,Funai:2018esm} or could even lead to instabilities (which come under the name of vanishing and exploding gradient~\cite{pascanu2013difficulty,DBLP:journals/corr/IoffeS15}). 
At the same time, hidden layers  with more units tend to overlearn specifics of the concrete (practically finite) data set they are exposed to, while overshadowing the typical traits of the given target system from which the training subset descends. This overlearning (also called overfitting) decreases the ability of our \textsc{ml} model to generalize the ``knowledge'' acquired during training about a target system  to new unseen data.

Evidently, the question arises about  optimal architectural choices that keep a balance between learning the desired features of the target data at a satisfactory level and overlearning irrelevant details from the training sample. A priori, 
efficiently fine-tuning such hyperparameters requires experience and a good understanding of the target system.  
At a more pragmatic level, to address this issue one usually scans over the hyperparameter space after imposing various constraints based on intuition and/or rules of thumb, in the spirit of e.g.~\cite{hinton2012practical}. 
Another practical approach~\cite{domhan2015speeding} is to train one simpler \textsc{ml} routine to detect the most optimal values for the  hyperparameters of the \textsc{ml} model which tries to learn the  target data.
At a more formal level, there exists mainly the widely used method of $\ell_p$\,--\,\textit{regularization}, where a \textsc{ml} model consisting of bigger hidden layer(s) is implemented that imposes a penalty for using a growing number of hidden resources~\cite{doi:10.1162/neco.1995.7.2.219}.
Despite its practical applicability, $\ell_1$\,--\, and $\ell_2$\,--\,regularization still requires a certain amount of fine-tuning to control the severity of  penalization for using additional hidden nodes and to adjust the consequent interference with training. 

\begin{figure}[t]
\begin{center}
\includegraphics[width=14cm,height=14cm,keepaspectratio]{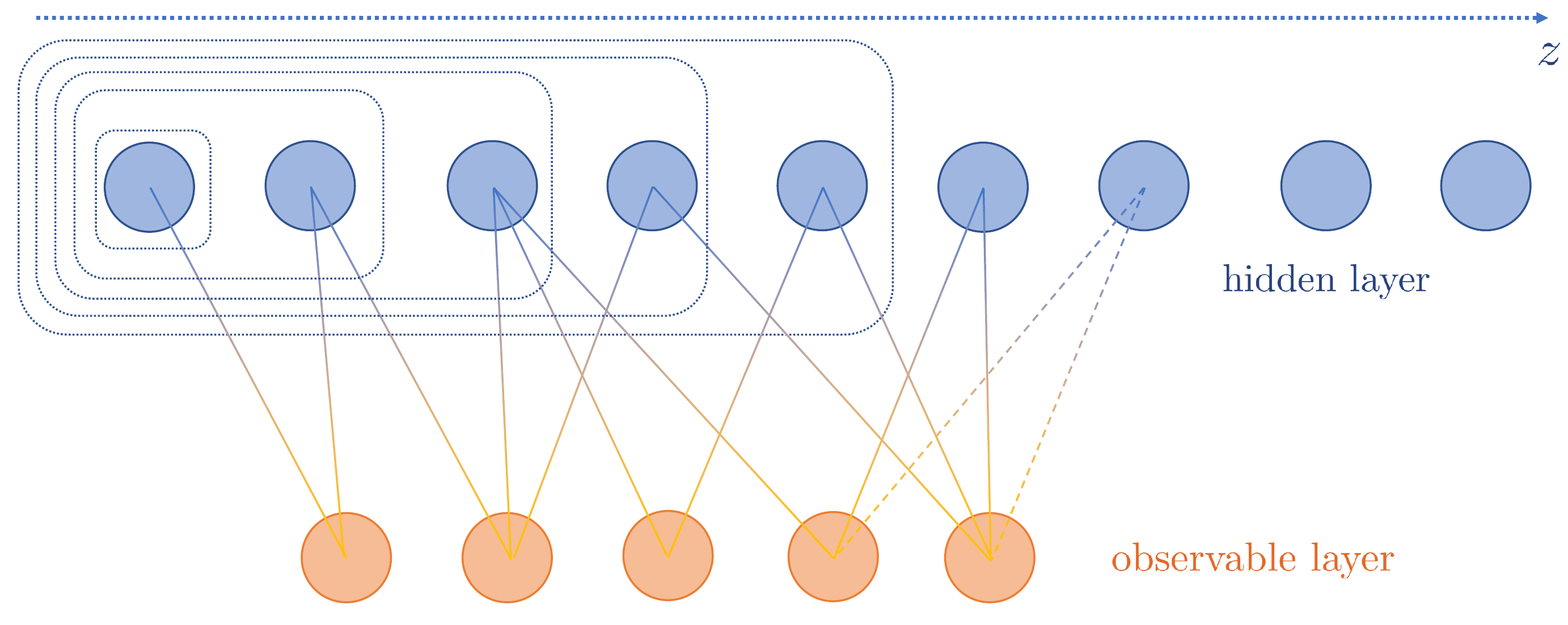}
\caption{The  concept of the grand-canonical extension to the \textsc{rbm}.
Hidden units are ordered from left to right
into layers of different lengths $z$ (dashed rectangles).  
Each hidden layer is  proportionally penalized according to $z$ by a chemical potential which is dynamically determined during training. Provided an observable distribution hidden layers of the appropriate size are invoked to model the target data.}\label{fig:vRBM_concept}
\end{center}
\end{figure}

In this paper, we aim at trying to eliminate the hyperparameter related to the number of hidden units altogether, in a dynamic 
fashion, i.e.\ as a solution to the extremization problem constituting the training procedure.
To this end, we concentrate on energy-based generative models which are trained to reproduce a target distribution by assigning a higher probability and lower energy to  physically occurring configurations (see~\cite{Goodfellow-et-al-2016} for a pedagogical introduction).
Specifically, the familiar restricted Boltzmann machine (\textsc{rbm}), originally formulated in~\cite{hinton1986learning} in the canonical ensemble, is reviewed and extended within the grand-canonical ensemble of statistical systems.  
Most naturally, this grand-canonical extension to accommodate a varying number of hidden units can be thought of as encompassing (theoretically infinitely) many restricted Boltzmann machines with hidden layers of all possible lengths. This concept is schematically presented in Figure~\ref{fig:vRBM_concept}.
Notice that  \rbm of various sizes $z$ that are used to model the target data share hidden units.

In the language of statistical mechanics, the theory is at finite chemical potential $\mu$, which now controls the strength of regularization, i.e.\ the most optimal size(s) of the hidden layer to be used.  
In principle, the \textsc{ml} model examined as a statistical system on its own right exhibits different phases depending on the value of $\mu$. 
By an appropriate choice of the  form of the chemical potential though, as a function of the 
other parameters that already exist in the Boltzmann machine (i.e.\ weights and biases),
its value  can be  dynamically determined during training to favour  networks of smaller sizes. 
In other words, the solution to the extremization problem posed during training in the grand-canonical formulation automatically ensures that our \textsc{ml} model learns the target distribution by promoting networks of the smallest possible hidden layer, avoiding thus overlearning.
In practice, we have to impose a cutoff to the maximal size of hidden layers that the grand-canonical  model could use.
For a sufficiently high cutoff though, not only we anticipate that the theory effectively becomes independent  of the concrete cutoff implemented, but furthermore that most designer choices concerning the precise  functional form of the chemical potential converge to the same regularizing effect. 

The method of regularization presented in this paper fundamentally differs from the familiar $\ell_p$\,--\,regularization w.r.t.\ two aspects.
On the one hand, once the chemical potential is appropriately chosen as a function of the rest of the \rbm parameters, there should remain no adjustable parameter --\,discrete or continuous\,-- related to the strength of regularization.  
On the other side, this regularization scheme naturally treats target data in a local fashion. 
This means that networks with a different number of hidden units will be invoked for  different  subsets of the target data depending on concrete features of each subset.
In the language of mathematical optimization, training an \rbm under  standard $\ell_p$\,--\,regularization poses a hardly constrained problem, while training the suggested grand-canonical extension with an appropriately chosen chemical potential results into a softly constrained problem.

\subsection*{Overview of the paper}

This paper is structured into a theoretical (Section~\ref{sc:Theory}) and applied (Section~\ref{sc:Training}) part.
Specifically, we review in Section~\ref{ssc:vRBM}  the necessary theoretical framework of grand-canonical Boltzmann distributions and lay out the model we wish to investigate. 
Subsequently, we set up in Section~\ref{ssc:CD} the stage for training,
by revisiting the minimization of the cross entropy between target and model distribution, while explaining necessary modifications at the level of a numerical solution. 
Next, we discuss in Section~\ref{ssc:ChemicalPotential} working assumptions and justify our concrete choice of the chemical potential as a function of the weights and biases that penalizes larger hidden layers.  
Ultimately, we apply the developed techniques to  two well-known data sets:  two-dimensional Ising configurations and the \textsc{mnist} set of handwritten digits, in Sections~\ref{ssc:Ising} and~\ref{ssc:MNIST}, respectively. We discuss and compare the learning outcome among the two paradigms as well as to the standard \rbm (without regularization).

\section{Varying number of hidden units}
\label{sc:Theory}

A restricted (i.e.\ in absence of intra-layer interactions) Boltzmann machine consists of an observable layer with $N$ units $\vset \equiv \{ v^i\}_{i=1,...,N}$
and a hidden layer with $z$ units $\hzset \equiv \{ h^a\}_{a=1,...,z}$. 
In this picture, the observed interactions among the units $\vset$ are modelled  via their connection  to the hidden (or latent) units $\hzset$.
Generically, Boltzmann machines are characterized by the weights (also called connections) $\weight_{ai}$ among observable and hidden layer together with the hidden and observable biases, $\bias_a$ and $\obias_i$ respectively.
In the following, we collectively denote the \textit{trainable} \rbm parameters by $\gl\equiv \{\weight_{ai},\bias_a,\obias_i\}$.
In contrast to those parameters which are expected to be fixed during training by extremizing the appropriate information-theoretic metric, the number of hidden units $z$ is a so-called \textit{hyperparameter} which needs to be fine-tuned beforehand.

In what follows, the question we are going to answer is how to eliminate this hyperparameter from training or in other words, leaving the size of the hidden layer unconstrained if and how the machine can ``select'' by itself optimum values for $z$ to explain the provided data. 
For brevity, we shall refer to a restricted Boltzmann machine invoking a varying number of hidden units as v\textsc{rbm}.

\subsection{Boltzmann machine at finite chemical potential}
\label{ssc:vRBM}

Henceforth, we focus for concreteness  on  a binary domain where $v^i=\pm1$ and Bernoulli Boltzmann machines also with  $h^a=\pm1$; the generalization of our discussion to Gaussian or other multimodal models being straight-forward.
As we are interested to work in this paper with varying number of hidden units $z\in \mathbb N$, we need to switch to the grand-canonical ensemble.
In this picture, the proper energy-based model at finite chemical potential $\mu$ is given by the grand-canonical Boltzmann 
distribution
\equ{
\label{GrandCanonical_BoltzmannDistribution}
p(\vset,\hzset,z) = \frac1Z \e^{-E-\mu\,z}
\quads{\text{with}}
E\equiv E(\vset,\hzset,z) = -\sum_{a=1}^z \sum_{i=1}^N h^a\weight_{ai} v^i  - \sum_{a=1}^z h^a\bias_a  - \sum_{i=1}^N \obias_i v^i
~,
}
%
$Z$ being the (generally intractable) partition function. 
Indices are contracted in Euclidean space and summations are explicitly recorded, for clarity. We work in units where $[k_BT]=1$.
Including  terms up to quadratic order in the units $v^i$ and $h^a$
helps efficiently solve the extremization problem posed during training, as outlined in Section~\ref{ssc:CD}.
Setting $\mu=0$ in Eq.~\eqref{GrandCanonical_BoltzmannDistribution} recovers the ordinary \rbm dictated by the canonical Boltzmann distribution.

Introducing the shorthand notations
\equ{
\sum_{\vset} \equiv \prod_{i=1}^N \sum_{v_i}
\quads{\text{and}}
\sum_{\hzset} \equiv \prod_{a=1}^z \sum_{h_a}
}
it is useful to define the free energies in the grand-canonical ensemble for a hidden layer of length $z$ via 
\begin{align}
\label{GrandCanonical:FreeEnergy1a}
F_{\gl,\mu}(\vset,z) =&\,\, 
\mu\,z - \sum_{i=1}^N \obias_i \,v^i - z\log2 - \sum_{a=1}^z \log\cosh\left(\sum_{i=1}^N \weight_{ai}\,v^i+\bias_a\right)
\\[0.7ex]
\label{GrandCanonical:FreeEnergy1b}
\text{and}\quad F_{\gl,\mu}(\hzset,z) =&\,\,
\mu\,z - \sum_{a=1}^z h^a\bias_a - N\log2 - \sum_{i=1}^N \log\cosh\left(\sum_{a=1}^z h^a\, \weight_{ai}+\obias_i\right)
\end{align}
by integrating out  hidden and observable variables, respectively.
$z$-independent terms like $N\log2$ can be dropped.
At most, we can absorb an irrelevant for our purposes (since it leads to a $z$-suppression irrespective of the training procedure)
term like $z\log2$ by redefining the chemical potential. 
%
%
Integrating out the hidden variables and summing over all possible lengths $z$ of the hidden layer leads to the observable free energy of the v\textsc{rbm},
\equ{
\label{GrandCanonical:FreeEnergy}
F_{\gl,\mu}(\vset)
= - \log\sum_{z=1}^\infty \e^{-F_{\gl,\mu}(\vset,z)} 
=
-\sum_{i=1}^N \obias_i v^i - \log \left[\sum_{z=1}^\infty \e^{-\mu\,z} \prod_{a=1}^z 2\cosh\left(\sum_{i=1}^N \weight_{ai}\, v^i +\bias_a\right)\right]
~,
}
in terms of which the  partition function of the model   compactly reads
\equ{
\label{vRBM:PartitionFunction}
Z\equiv Z(\gl,\mu) =
\sum_{\vset} \e^{-F_{\gl,\mu}(\vset)}
~.
}
Evidently, the associated probabilities we are going to use in the following section are generically given by 
\equ{
\label{vRBM:ModelProbabilties}
p_{\gl,\mu}(\vset) = \frac{1}{Z}
e^{-F_{\gl,\mu}(\vset)}
\qand 
p(\vset,z) = \frac{1}{Z}e^{-F_{\gl,\mu}(\vset,z)}
~.
}

%
At this stage,
the partition function $Z$ of the \vrbm depends on the ordinary \rbm parameters $\gl$ as well as the chemical potential $\mu$.
This model can be viewed as a collection of ordinary Boltzmann machines with a hidden layer of different  lengths $z$.
Hence, \rbm with different number of hidden units contribute to grand-canonical expectation values
weighted by the introduced chemical-potential term. 
From the point of view of statistical systems, our \vrbm is expected to exhibit different phases: For $\mu\gg1$ smaller hidden layers are favoured prohibiting  learning the desired distribution, while in the opposite limit networks with larger number of hidden units prevail and overfitting occurs. 
%

%
\paragraph{Working assumptions.}

For the intended \textsc{ml} application, we can only sum over a finite number $K$ of sizes $z$ of the hidden layer. We thus have to use  the  free energy 
\equ{
\label{GrandCanonical:FreeEnergy_approx}
F_{\gl,\mu,K}(\vset)
= - \log\sum_{z=1}^K \e^{-F_{\gl,\mu}(\vset,z)} 
~,
}
whose limiting case is Eq.~\eqref{GrandCanonical:FreeEnergy}.
This model has one more parameter than the 
partition function Eq.~\eqref{vRBM:PartitionFunction} 
of the idealized v\textsc{rbm}: the maximum number $K$ of hidden units   that the model has at its disposal. 
In the spirit of eliminating hyperparameters and assuming that only a finite number of hidden units is needed to explain the target data, we take the number of available hidden units to be very large  until the model ``thinks'' it always has a sufficient number of 
resources.
Formally, we shall work in the regime of \textit{large} $K$. To quantify this, $K$ needs to be sufficiently larger than the natural scale of the problem set by the number of observable units $N$, so that effects induced by the finite amount of hidden resources are suppressed by powers of $N/K\ll 1$. 
%
In the following, we shall always work in the  approximation $F_{\gl,\mu,K\gg N}(\vset) \approx F_{\gl,\mu}(\vset)$.
In practice, as we are going to see in Section~\ref{sc:Training} this can be relaxed to $K>N$. 

Even under the large-$K$ regime, the model described by partition function Eq.~\eqref{vRBM:PartitionFunction} still seems to have annoyingly many adjustable parameters, for sure not less than its canonical counterpart.
Most crucially, we are not interested to merely swap the discrete number $z$ of hidden units with a continuous chemical potential $\mu$, but to eliminate the hyper-parameter determining the size of the hidden layer from the training altogether. Hence, we shall take the chemical potential to be some function of the other \rbm parameters, $\mu\equiv\mu(\gl)$. 
Generically, there are a few desired properties this function is expected to possess.
First of all, at the formal level, the chemical potential being an intensive variable will be treated as a global model parameter, i.e.\ it will not exhibit any explicit or implicit dependance on the hidden-layer number $z$. 
At the practical level, dropping this constraint\footnote{For such a treatment of \vrbm with a local $z$-dependent chemical potential see~\cite{2015arXiv150202476C,nalisnick2017infinite,peng2018better}.}  leads to certain instabilities in the learning process, at least for the systems tested in this work.
In addition, to ensure that $\mu$ has a regularizing effect at all, it should be independent from the sign of the other parameters, weights $\weight$ and biases $\bias$.
Given those formal requirements one is entitled to test which ansatz for $\mu$ best suits the training on a particular target system. 

%
For the purposes of this paper, we adopt a full agnostic approach concerning the target system and refrain from imposing any designer biases. Specifically, as the biases $\obias_i$ solely characterize observable units, the chemical potential $\mu$ --\,being a feature that is intimately related to the length of the hidden layer\,-- is not expected to {explicitly} depend upon those.  
In the same spirit, we further assume that 
$\mu$ is unbiased towards observable units, i.e.\ it does not exhibit any {implicit} dependance on them, as well. 
Thus, our chemical potential is at most a function of the row p-norm of the weight matrix and hidden biases,
%
\equ{
\label{ChemicalPotential_FunctinalForm}
\mu\equiv\mu\left(\sum_{i=1}^N \abs{\weight_{ai}}^p, \bias_a\right)
~.
}
In what follows, we take all model quantities to ultimately depend only on the trainable parameters $\gl$ that are determined by appropriate  training, as outlined in the next paragraph. 


\paragraph{The extremization condition.}
A generative model is primarily used to learn to reproduce a given data distribution $q(\vset)$ by extracting its characteristic features. In the following, we always take the domain of $q(\vset)$ to coincide with the domain of both the observable and hidden units of our generative model.
Training our Boltzmann machine on a given {target} probability $q(\vset)$ is performed by extremizing w.r.t.\ model parameters $\gl$ (recall that we take $\mu\equiv\mu(\gl)$) an appropriately chosen {target} function. 
As a target function we conventionally take the opposite of the cross entropy between model and target distribution, i.e.\
the expectation value under target distribution $q(\vset)$  of the logarithm of probability $p_\gl(\vset)$ given in Eq.~\eqref{vRBM:ModelProbabilties}:
\equ{
t(\gl) \equiv \sum_\vset q(\vset) \log p_\gl(\vset)
~.
}
It is straight-forward to show that maximizing $t(\gl)$  is equivalent to minimizing the Kullback-Leibler divergence which gives the relative entropy between target $q(\vset)$ and model $p_\gl(\vset)$ distribution.
%

%
First, note that deriving free energy Eq.~\eqref{GrandCanonical:FreeEnergy} w.r.t.\ some trainable parameter $\gl$ we get 
\equ{
\label{vRBM:DerivativeOfFreeEnergy}
\frac{\partial }{\partial \gl} F_\gl(\textbf v)
= \sum_{z=1}^K p_\gl\left(z\vert\textbf v\right) \frac{\partial F_\gl(\textbf v,z)}{\partial\gl}
~,
} 
where the conditional probability associated to Eq.~\eqref{GrandCanonical:FreeEnergy1a} reads
\equ{
\label{z_conditionalProbability}
p_\gl\left(z\vert\vset\right)
=
\frac{p_\gl\left(\vset,z\right)}{p_\gl\left(\vset\right)} = 
\frac{\e^{-F_\gl(\vset,z)}}{\e^{-F_\gl(\vset)}}
=
\frac{e^{-\mu z}\prod_{a=1}^z 2\cosh \left(\sum_{i=1}^N \weight_{ai}\,v^i+\bias_a\right)}{\sum_{z'=1}^Ke^{-\mu z'}\prod_{a'=1}^{z'} 2\cosh \left(\sum_{i'=1}^N \weight_{a'i'}\,v^{i'}+\bias_{a'}\right)}
~.
}
Next, we can express the extremization condition on $t(\gl)$ in terms of a derivative of the free energy Eq.~\eqref{GrandCanonical:FreeEnergy1a}
at level $z$:
%
%
\begin{align}
\label{ExtremizationCondition}
0\overset{!}{=}
\frac{\partial }{\partial \gl} t(\gl)
=&\,\,
\sum_{\vset}  \sum_{z=1}^K \left[q\left(\textbf v\right) p_\gl\left(z\vert\textbf v\right) - p_\gl\left(\textbf v,z\right) \right]\frac{\partial (-F_\gl(\textbf v,z))}{\partial\gl}
~.
\end{align}
%
%
Using the concrete form~\eqref{GrandCanonical_BoltzmannDistribution} of grand-canonical Boltzmann distribution the extremization condition~\eqref{ExtremizationCondition} results into a set of $(K\times N+K+N)$ equations, 
\begin{align}
\label{ExtremizationCondition_Explicit}
\sum_{\textbf v}& \sum_{z=1}^K 
\left\lbrace 
H(z-a)  
\tanh\left(\sum_{i'=1}^N \weight_{ai'} v^{i'} + \bias_a\right) v^i
-\frac{\partial\mu}{\weight_{ai}} \, z
\right\rbrace
\left[q\left(\vset\right) p_\gl\left(z\vert\vset\right) - p_\gl\left(\vset,z\right) \right]
= 0
\nonumber
\\[1ex]
\sum_\vset & \sum_{z=1}^K\left\lbrace 
H(z-a) \tanh\left(\sum_{i'=1}^N \weight_{ai'} v^{i'} + \bias_a\right)
-\frac{\partial\mu}{\bias_a} \, z
\right\rbrace
\left[q\left(\vset\right) p_\gl\left(z\vert\vset\right) - p_\gl\left(\vset, z\right) \right]
=  0
\nonumber
\\[1ex]
\sum_\vset & v^i
\left[q\left(\vset\right)  - p_\gl\left(\vset\right) \right] =  0
~,
\end{align}
taking $\gl=\weight_{ai}, \bias_a, \obias_i$ respectively.
Notice in particular, 
the appearance of Heaviside step function $H$ with $H(x)=1$ for $x\geq0$, as a consequence of varying  number of hidden units as well as the additional term in the first two equations due to the derivative of chemical potential Eq.~\eqref{ChemicalPotential_FunctinalForm}.
%

%
In the literature, such extremization conditions are often compactly written as 
\begin{align}
\label{ExtremizationCondition_Compact}
\left\langle h^a v^i \right\rangle_\text{data} - \left\langle h^a v^i \right\rangle_\text{model}
-
\frac{\partial\mu}{\weight_{ai}}\left(\VEV{z}_\text{data} - \VEV{z}_\text{model}\right) =&\,\, 0
\nonumber
\\[0.7ex]
\langle h^a\rangle_\text{data} - \langle h^a\rangle_\text{model}
-
\frac{\partial\mu}{\bias_a}\left(\VEV{z}_\text{data} - \VEV{z}_\text{model}\right) =&\,\, 0
\nonumber
\\[0.7ex]
\langle v^i\rangle_\text{data} - \langle v^i\rangle_\text{model} =&\,\, 0
~,
\end{align}
where expectation values are understood to be taken w.r.t.\ the distribution of the provided training data $q(\vset)$ and the probability distribution $p_\gl(\vset)$ generated by our model. 
Due to our inability to generically evaluate the partition function $Z(\gl)$ the derived set of conditions cannot be solved in any closed form. 

\subsection{Contrastive Divergence revisited.}
\label{ssc:CD}
To circumvent this we shall make a gradient-descend inspired approximation. 
Employing \textit{contrastive divergence} (\textsc{cd}) has proven to be a particularly efficient way 
to numerically find the maximum, Eq.~\eqref{ExtremizationCondition}, by updating our model parameters $\gl$ in the direction of steepest descend according to
\equ{
\label{CD:updateStep}
\gl^{(\ga+1)} = \gl^{(\ga)} + \gamma\, \nabla_\gl^{(\ga)}
\VEV{\log p(\vset)}_\text{data}
}
until convergence is achieved. 
The \textit{learning rate}  $\gamma$  is a tuneable parameter of the extremization process itself. Too large values of $\gamma$ could drive away from the desired extremum, while too small learning rates slow down the training process. 
The \textsc{cd} derivative at $\ga$-th step is defined by 
\equ{
\nabla_\gl^{(\ga)} \equiv \beta\, \nabla_\gl^{(\ga-1)} + \left(1-\beta\right)\frac{\partial}{\partial\gl^{(\ga)}} 
~,
}
where $\ga=0,1,...$ with $\nabla_\gl^{(-1)}=0$. The \textit{momentum} $\beta\in[0,1)$ acts as a ``memory'' of previous updates to ensure stability of the gradient-descent algorithm. 

\paragraph{The mean-field ansatz.}
For  each \textsc{cd} update in Eq.~\eqref{CD:updateStep} we need to compute 
the expectation values in Eq.~\eqref{ExtremizationCondition_Compact}
that appear when taking the derivative of the target function $t(\gl)$ w.r.t.\ model parameters $\weight_{ai}$, $\bias_{a}$ and $\obias_i$.
For that, we use mean-field theory to iteratively compute 
the \vev of $v^i$, $h^a$ and $z$ using consistency equations and substituting the one into the other.
In our \vrbm setting though, 
a slight modification of the mean-field theoretic consistency conditions which are invoked by the \textsc{cd} method in the standard \textsc{rbm} construction is required.
Schematically, starting from the distribution $\vset$ of observable units
the length of hidden layer $z$ and subsequently the hidden units $\hzset$ are to be inferred which in turn are about to give a new estimation for $\vset$.

In detail, provided the distribution of observable units $q(\vset)$
it is straight-forward to compute the multimodal conditional probability $p(z\vert\vset)$ given in Eq.~\eqref{z_conditionalProbability}. 
As in ordinary Boltzmann machines, we calculate the expectation value of the hidden units provided a sample $\vset$ from the observable distribution
from the (grand-canonical in our case) free energy Eq.~\eqref{GrandCanonical:FreeEnergy}  by 
\equ{
\label{h_VEV}
\langle h^a \rangle_{\vset}
=
\frac{\partial}{\partial \bias_a} \left(-F(\vset)\right)
=
\sum_{z=1}^K \frac{\partial(-F(\vset,z))}{\partial \bias_a} p(z \vert \vset) 
=
P(z\geq a\vert \vset)\tanh \left(\sum_{i=1}^N \weight_{ai}\, v^i + \bias_a\right)
}
using Eq.~\eqref{vRBM:DerivativeOfFreeEnergy}.
The complementary cumulative distribution function (ccdf) or survival function 
of the probability $p(z\vert\vset)$, 
\equ{
\label{ccdf}
P(z\geq a\vert \vset) \equiv \sum_{z=a}^K p(z \vert \vset)
~,
} 
ensures that only layers including the $a$-th hidden unit contribute to its  conditional expectation value, cf.\ Eq.~\eqref{ExtremizationCondition_Explicit}.
Next, we need to determine some optimum value for the size of the hidden layer $z$ being sufficient to extract the features from the provided observable distribution.
In principle, the number $z$ of hidden units has to be  determined 
by sampling from $p(z\vert\vset)$.
In practice, it is computationally cheaper while converging faster to either compute the \vev of $z$ provided $\vset$,
\equ{
\label{z_conditionalVEV}
\VEV{z}_\vset 
=
\sum_{z=1}^K z\, p(z\vert\vset)
~,
}
and round it upwards to the nearest integer,
or take the largest most probable value for $z$ given its conditional distribution,
\equ{
\label{z_probableValue}
z_\text{probable} = \max\left\lbrace z ~\vert~ \max_z \,p(z\vert \vset)\right\rbrace
~.
}
After sufficient number of training epochs and at  large $K$,  all methods effectively lead to the same outcome.
%
%
%
%
%
Ultimately, given the derived value of $z$ together with $\VEV{h^a}_\vset$ 
we are in a position to deduce a new estimation for $v^i$. Such an estimation has to be extracted from a Boltzmann machine with $z$ hidden units taking values $\VEV{h^a}_\vset$
via
\begin{align}
\label{v_VEV}
\langle v^i \rangle_{\hzset,z} =
\frac{\partial(-F(\hzset,z))}{\partial \obias_i}
=
\sum_{v^i} v^i\, p(v^i \vert \hzset,z)
=
\tanh \left(\sum_{a=1}^z h^a\weight_{ai} +\obias_i \right) 
~,
\end{align}
where the necessary conditional probability is deduced from 
Eq.~\eqref{GrandCanonical:FreeEnergy1b},

\equ{
\label{probability_v_conditioned_h_z}
p(v^i\vert \hzset,z) = 
\frac{p(v^i,\hzset,z)}{p(\hzset,z)}
=
\frac{\e^{-E(v^i,\hzset,z)-\mu z}}{\e^{-F(\hzset,z)}}
=
\frac{\e^{\left(\sum_{a=1}^z h^a\weight_{ai}+\obias_i\right)v^i}}{2 \cosh \left(\sum_{a=1}^z h^a\weight_{ai}+\obias_i\right)}
~.
}

Hence, $z$ hidden units contribute to the feature extraction from the desired data set leading in turn, to the generation of a new estimation for $\{v_i\}$.   
As our estimation for $z$, either Eq.~\eqref{z_conditionalVEV} or Eq.~\eqref{z_probableValue},  depends on the initial configuration $\vset$ that we sample from input distribution $q(\vset)$, we observe that \textit{different} configurations from $q(\vset)$ would generically be explained by a \textit{different} number of hidden units.
In other words, the \vrbm model has the freedom to adjust the size $z$ of its hidden layer depending on the complexity level of \textit{each} subset in $q(\vset)$.
This observation constitutes one of the fundamental departures from standard $\ell_p$\,--\,regularization~\cite{bishop2006pattern} (the other main difference being the absence of a continuous parameter conventionally controlling the strength of ordinary regularization schemes).
When regularization is globally applied to the input set $q(\vset)$ the local features of each example configuration $\vset$ are detected by the same fixed number of hidden units encompassing the danger of over-learning for some of the subsets in $q(\vset)$.

In total, the $k$-th iteration in  mean-field theory for $k\in\mathbb N$
reads
\equ{
\label{kcd_step}
z_{(k-1)}\equiv \VEV{z}_{\vset^{(k-1)}}
~\text{ and }~
h^a_{(k-1)} \equiv \VEV{h^a}_{\vset^{(k-1)}}
\quads{\longrightarrow}
v^i_{(k)} \equiv \VEV{v^i}_{\hset^{(k-1)},\,z^{(k-1)}}
~,
}
under the initial configuration $v^i_{(0)}\equiv v^i$ described by $q(\vset)$.
For well-defined extrema,  mean-field theory is formally expected to converge towards $\VEV{v^i}$, the very latest as $k\rightarrow\infty$. In practice, the mean-field ansatz converges for physical data  beyond numerical accuracy already when $k=2$.
%

%
%
\paragraph{Summary of the numerical method.}
All in all, putting everything together the $\ga$-th step of the \textsc{cd} method in the \vrbm dictated by Eq.~\eqref{CD:updateStep} under a mean-field theoretic ansatz after $k$ steps in Eq.~\eqref{kcd_step}  consists of 
\begin{align}
\label{a_CDupdate}
\weight^{(\ga+1)}_{ai} =&\,\, \weight^{(\ga)}_{ai} 
+ \gamma \left\lbrace\beta + 
\left(1-\beta\right)\delta_{ab}\,\delta_{ij}\left[ {h^b_{(0)}} {v^j_{(0)}} - {h^b_{(k)}}{v^j_{(k)}} -\frac{\partial\mu}{\partial \weight^{(\ga)}_{bj}} \left(z_{(0)}-z_{(k)} \right)\right] \right\rbrace
\nonumber
\\[0.7ex]
\bias^{(\ga+1)}_{a} =&\,\, b^{(\ga)}_{a\phantom{ii}} 
+ \gamma \left\lbrace\beta + 
\left(1-\beta\right)\delta_{ab}\left[ {h^b_{(0)}} - {h^b_{(k)}} -\frac{\partial\mu}{\partial \bias_{b}^{(\ga)}} \left(z_{(0)}-z_{(k)} \right)\right] \right\rbrace
\nonumber
\\[0.9ex]
\obias^{(\ga+1)}_{i} =&\,\, \obias^{(\ga)}_{i\phantom{ai}} 
+ \gamma \left\lbrace\beta + 
\left(1-\beta\right)\delta_{ij}\left[ {v^j_{(0)}} - {v^j_{(k)}}\right] \right\rbrace
~.
\end{align}
Kronecker delta is used to raise and lower indices. 
Compared to the standard \rbm its grand-canonical extension adds a complexity in computing conditional probability Eq.~\eqref{z_conditionalProbability} which grows linearly in the maximum number of hidden units $K$ the model has at its disposal, cf. Eq.~\eqref{GrandCanonical:FreeEnergy_approx}.
Apart from computing this multimodal distribution, the grand-canonical \textsc{cd} algorithm shares everything with its canonical counterpart.
Setting $\mu=0$ in the equations~\eqref{a_CDupdate} and taking $p(z\vert \vset)=\delta_{z,z_0}$ in Eq.~\eqref{h_VEV}\,--\,\eqref{probability_v_conditioned_h_z} reproduces the ordinary \textsc{cd} algorithm of the \rbm in the canonical ensemble with fixed number $z_0$ of hidden units.

\subsection{A penalizing chemical potential}
\label{ssc:ChemicalPotential}

So far, we have avoided to concretely specify the form of the chemical potential, besides a more generic discussion towards the end of Section~\ref{ssc:vRBM}.
In particular, we have explained our motivation in taking $\mu\equiv\mu(\gl)$ and discussed about the anticipated functional form of $\mu>0$ concluding to Eq.~\eqref{ChemicalPotential_FunctinalForm} in order to regulate the length of the hidden layer.
Of course, being an intensive quantity $\mu$ could well implicitly or explicitly depend on global \textit{non}-trainable parameters like $N$ and $K$, but not on $z$ itself. 

Given that the additional term in the grand-canonical ensemble is expected to naturally act as a \textit{regularizer} for the number of hidden units and by naive dimensional analysis, the chemical potential should have the form of a \textit{non}-negative energy density:
\equ{
\label{ChemicalPotential:ConceptualAnsatz}
\mu = \frac{\text{non-negative ``vacuum energy''}}{\text{number of active hidden units}}
~,
}
where the
{number of active hidden units equals the number of $h^a$ participating in explaining the target data, i.e.\ have non-vanishing weights $\weight_{ai}$ for at least one $i$. 
%
Conceptually, our ansatz~\eqref{ChemicalPotential:ConceptualAnsatz} for the chemical potential describes some notion of total non-negative --\,to achieve a regularizing effect\,-- vacuum energy (which could well be infinite  when $K\rightarrow\infty$) that is equally partitioned 
over each hidden unit actively participating in extracting the features of the target data.
That way, we uniformly define a penalty the model needs to pay for using additional hidden units.
Networks 
with larger hidden layers are then proportionally penalized to their length $z$ by a factor $\mu$ that equals the aforementioned ``vacuum-energy'' density.

Even under the working assumptions of paragraph~\ref{ssc:vRBM}, there is still a certain freedom in specifying the form of the total ``vacuum energy'' in Eq.~\eqref{ChemicalPotential:ConceptualAnsatz}.
Inspired mainly from the regularization procedure performed~\cite{hinton2012practical,wang2017regularization} in ordinary \rbm
it seems plausible to take as a definition of the ``vacuum energy'' entering the chemical potential,
\equ{
\label{VacuumEnergyDEF}
\text{non-negative ``vacuum energy''} = 
\sum_{a=1}^K\feq_a
~,
}
introducing the fundamental ``vacuum-energy'' quantum 
%
\equ{
\label{FundamentalEnergyQuantum}
\feq_a = 
\sfrac1N\sum_{i=1}^N \abs{\weight_{ai}}^p +  \abs{\bias_a}^p 
}
characterizing each hidden unit.
Notice that $\feq_a$ depends on the $a$-th row $p$\,--\,norm of the weight matrix and the respective hidden bias. 
In a similar spirit, the matrix norm of $\omega$ appearing in definition~\eqref{VacuumEnergyDEF} avoids  placing any bias  over some specific observable $v^i$ or hidden $h^a$ unit that could falsify a fair  learning of the necessary connections $\omega_{ai}$.

Concerning the denominator of Eq.~\eqref{ChemicalPotential:ConceptualAnsatz}
that should count the number of hidden units actively used by the \vrbm to model the given data
it is straight-forward to approximate it by a smooth (since $\tanh \feq_a \rightarrow1$ asymptotically when $\feq_a\gg1$) counter,
\equ{
K_\text{eff} = \text{number of active hidden units} \approx 
\sum_{a=1}^K\tanh \feq^a 
~.
}
In fact, to precisely count the number of active units we would have to replace $\tanh$ by \textsc{ReLu} function. However, the infinitely steep derivative of the latter function when its argument becomes zero renders the numerical solution presented in Section~\ref{ssc:CD} inadequate, as the \textsc{cd} method gets stack either to the $\mu\approx0$ regime or $\mu\gg1$ depending on initial conditions.
Hence in total, a candidate chemical potential depending only on the row-norm of $\weight$ and hidden biases $\bias$ which is unbiased towards any hidden or observable unit reads 
\equ{
\label{ChemicalPotential:ConcreteAnsatz}
\mu \equiv \mu(\weight,\bias) 
= \frac{\sum_{a=1}^K \left( \sfrac1N\sum_{i=1}^N \abs{\weight_{ai}}^p
+\abs{\bias_a}^p \right)}
{\sum_{a'=1}^{K}\tanh \left( \sfrac1N\sum_{i'=1}^N \abs{\weight_{a'i'}}^p+\abs{\bias_{a'}}^p \right)}
\,\equiv\,
\frac{\sum_{a=1}^K\feq_a}{\sum_{a'=1}^K\tanh\feq_{a'}}
~.
}
Alternatively, depending on the input data to ensure numerical stability one could further smoothen this definition 
by taking the $\cosh$ of Eq.~\eqref{VacuumEnergyDEF}. 
\begin{figure}[t]
\begin{center}
\includegraphics[width=16cm,height=16cm,keepaspectratio]{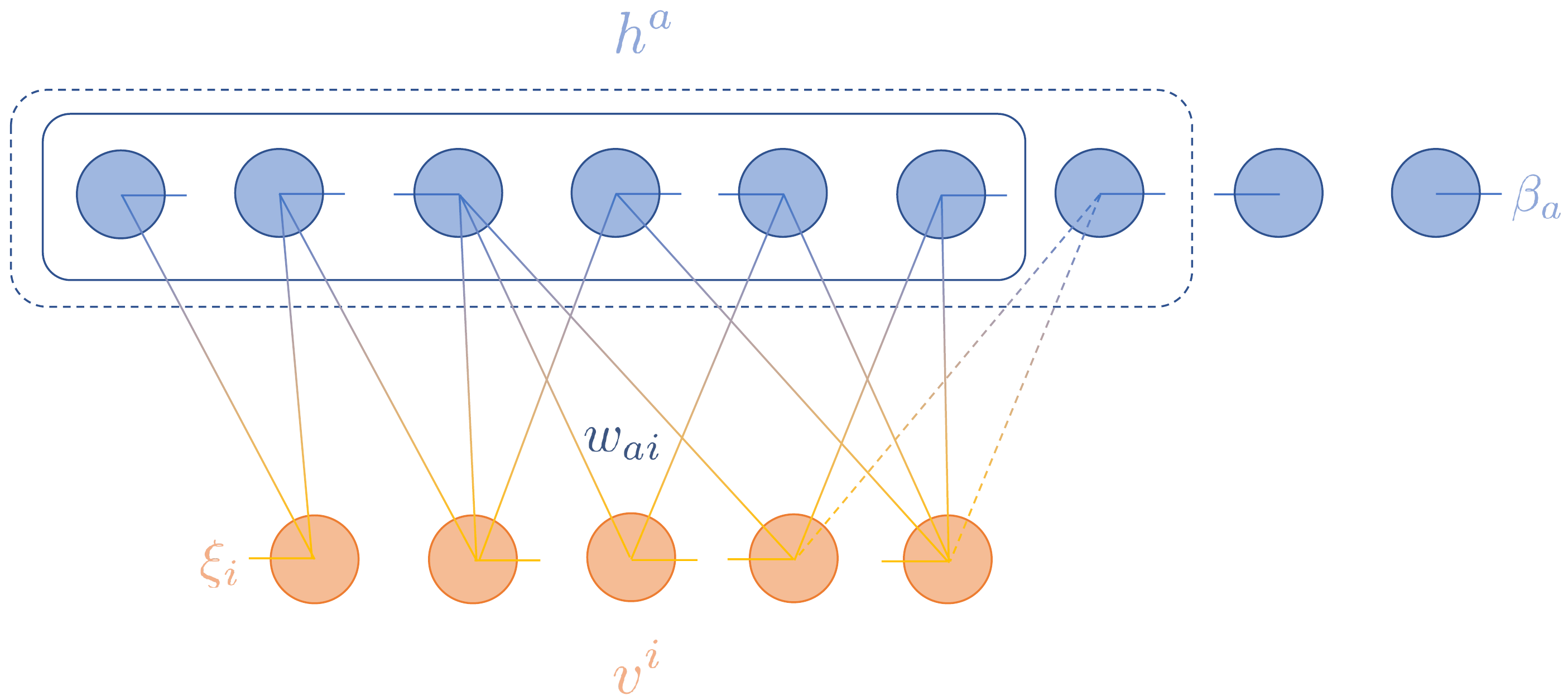}
\caption{In the self-regularizing \rbm the interactions among observable units $v^i$ with biases $\obias_i$ are  modelled by connecting via weights $w_{ai}$ to hidden units $h^a$.
To explain an observable configuration some of the  hidden units might not be used in spite of having weights (dashed lines) which have been trained on the full data set.
Additionally, there are hidden units which
do not participate in feature extraction, at all, while their biases $\bias_a$ still contribute to the regularization procedure.  
}\label{fig:vRBM_schema}
\end{center}
\end{figure}

At this point, a few remarks are in order:
Evidently, our energy quantum $\feq_a\geq0$ 
is solely controlled by the respective bias, when the hidden unit is inactive in the sense of $\weight_{ai}=0$ for all $i=1,..,N$.
Consequently, the model has the ability to recursively adjust $K_\text{eff}$ and $\mu$ by merely administrating the hidden biases $\bias_c$ (and hence $\feq_c$) of 
the latter $(K-K_\text{eff})$ inactive units $h^c$ without interfering much with the actual training of the first $K_\text{eff}$ units $h^b$.
To control the aforementioned interference between regularization and learning those first 
$h^b$ units the $b$-th row $p$-norm of the weight-matrix $\weight$ entering definition~\eqref{FundamentalEnergyQuantum} was   
normalized by the observable scale $N$.
For finite values of $K$, this seems to be the most sensible normalization  in order to avoid naively introducing any hierarchy in $N$ among the two summands in Eq.~\eqref{FundamentalEnergyQuantum}. 
At large $K$ (i.e.\ $K\gg N$) however, we anticipate (and indeed find) that  any \textit{generic} number in front of $\abs{\weight_{ai}}$ leads to the same outcome after sufficient number of epochs.
The same observation holds for any generic rescaling of the hidden biases entering Eq.~\eqref{FundamentalEnergyQuantum}. Despite that such redefinitions of the chemical potential could delay the convergence of the training algorithm, they  do not crucially change the regularizing effect of Eq.~\eqref{ChemicalPotential:ConcreteAnsatz} which is determined by our conceptual choice of the functional form Eq.~\eqref{ChemicalPotential:ConceptualAnsatz}.  

In Figure~\ref{fig:vRBM_schema}, we schematically draw how the \vrbm is expected to model interactions among five observable units $v^i$ by using their connection (continuous lines) to hidden units $h^b$ via the trained weights $w_{bi}$. 
Not all weights that the \vrbm has learned during training are expected to participate in the feature detection of each example from the target distribution $q(\vset)$.
In our schematic depiction, one  hidden unit remains inactive (dashed connecting lines) for the given observable configuration $\vset$ as dictated by $p(z\vert\vset)$.
The size of the biggest hidden network (wrapped by a dashed rectangular) encompassing all units $h^b$ connecting to the observable layer determines $K_\text{eff}$. 
Finally, there should exist hidden units $h^c$ for $c>K_\text{eff}$ whose weights are effectively regularized to zero. In fact, weights $w_{ci}$ with $c>K_\text{eff}$ quickly get exponentially suppressed by the chemical-potential term in $p(z\vert\vset)$  beyond  any meaningful numerical accuracy.
The role of those hidden units is to ensure the regularizing effect of the \vrbm through their biases $\beta_c$, as we are going to explicitly verify in the following  two sections. 

To summarize, the grand-canonical extension of the \rbm is expected  to dynamically determine during training the most optimal value of a penalizing  chemical potential $\mu$ in order to extract the typical characteristics of the  target distribution at a satisfactory level while avoiding over-learning.
To do so, it employs --\, provided a sufficiently large number of hidden units\,-- hidden layers of different lengths $z$  depending on the input configuration $\vset$  with a probability $p(z\vert \vset)$.


\section{Training Boltzmann machines at finite chemical potential}
\label{sc:Training}

In this section, we aim at applying the grand-canonical extension of the \rbm developed in the previous paragraphs to learn target distributions $q(\vset)$  that act as a reference in their respective fields.
For this purpose, two error functions which measure the deviation of the \textsc{ml}-generated data from the actual data set come in handy.
So far, the prediction of our model derived in Section~\ref{ssc:CD}} is an expectation value $\VEV{v^i}$ of the mean-field theoretic ansatz Eq.~\eqref{kcd_step} for each observable unit.
A priori, $\VEV{v^i}$ is not expected to  belong to the domain of $q(\vset)$, though. 
As those error measures are concerned  with the crucial ability  
of the generative model to learn and reproduce the target data, one should stochastically replace\footnote{In information-theoretic context and the computer-scientific literature~\cite{MacKay2003}, this procedure appears as Gibbs sampling.} the expectation value $\VEV{v^i}$  with the actual value $v^i$ 
as sampled from distribution~\eqref{probability_v_conditioned_h_z}.
For binary distributions, given the expectation value 
that is computationally more efficient to obtain,  
this replacement is simply done with a probability $P(v^i=\pm1)=(1\pm \VEV{v^i})/{2}$.

\paragraph{Loss functions.}
In most applications of interest, training the \textsc{ml} model on the full distribution $q(\vset)$, which is usually not fully known or very expensive to compute, is not a feasible task.
In practice, we train our generative model on a smaller number of 
$\mathcal N_\text{train}$ 
selected points $\vset_{A}=\{v_{A}^i\}_{i=1,...,N}$ sampled from $q(\vset)$.
For ease of notation, we summarize the training subset via 
\equ{
q_\text{train}(\vset) \equiv \frac{1}{\mathcal N_\text{train}}\sum_{A} \prod_{i=1}^N\delta(v^i-v^i_{A})
~,
}
in terms of one-dimensional delta functions (or Kronecker delta for discrete distributions).
To express the ability of the generative model to accurately learn to reproduce  
$q_\text{train}(\vset)$
we introduce the quadratic reconstruction error on training data (also called train loss function) by 
\equ{
\label{TrainErrorDEF}
\varepsilon_\text{train} :=
\sum_\vset q_\text{train}(\vset)
\sum_{i=1}^N
\left( v^i - \VEV{v^i} \right)^2
~.
}
For the expectation value we substitute the prediction 
$\vset^{(k)}$
of our model after $k$ steps in mean-field theory according to Eq.~\eqref{kcd_step} to closely follow the convergence of our algorithms.
%
%
To benchmark the quality of learning, i.e.\ to which extend our generative model has correctly identified important features of the target distribution to faithfully reconstruct new --\,unseen during training\,-- data points $\vset_B=\{v_B^i\}_{i=1,...,N}$ from $q(\vset)$, summarized by 
\equ{
q_\text{test}(\vset) \equiv \frac{1}{\mathcal N_\text{test}}\sum_{B} \prod_{i=1}^N\delta\left(v^i-v^i_B\right)
~,
}
we invoke the quadratic reconstruction error on test data (or test loss function)
\equ{
\label{TestErrorDEF}
\varepsilon_\text{test} :=
\sum_\vset q_\text{test}(\vset)
\sum_{i=1}^N
\left( v^i - \VEV{v^i} \right)^2
~.
}
Also here, we substitute 
our estimation $\vset^{(k)}$ of the 
expectation value  deduced by iteratively applying Eq.~\eqref{kcd_step} during training to monitor the learning quality.

In the same spirit, one could define absolute learning errors by taking the absolute value of the difference between target data and  mean-field theoretic outcome. 
This makes sense especially for continuous distributions (where the quadratic loss function could underestimate the  learning error in the domain $[0,1]$) or when outliers in the training data being overweighted by the quadratic loss erroneously lead to big training but still reasonable test errors. 
%
%
Evidently, $\varepsilon_\text{train} \rightarrow 0$ means that our \textsc{ml} algorithm is able to accurately reproduce the provided points from  target distribution $q(\vset)$,
while $\varepsilon_\text{test} \rightarrow 0$ signifies the ability of our model to generalize into unseen data after correctly extracting the characteristic traits of $q(\vset)$ during training. 

At first sight, the outlined way to use these loss functions 
seems to depart from their general objective, as described in the beginning, which is to judge the actual prediction on the domain of the generative model (i.e.\ the integer value $v^i$ for discrete distributions). 
However, after a sufficient number of epochs for the binary systems under consideration,
we observe that $\VEV{v^i}$ indeed converges towards the domain values $\pm1$.
Consequently
the probability $P(v^i=\pm1)$ to find the $i$-th unit with value $\pm1$ sharply peaks at 0 or 1 depending on the expectation value  $\VEV{v^i}$. In turn, this signals  that $\VEV{v^i}$ effectively coincides with the sampled value $v^i$ within the domain of our model.
For this reason, after a desired accuracy has been achieved, it is allowed to simply round $\VEV{v^i}$ to the nearest integer.
As we are primarily   interested in this work in benchmarking the convergence rate and learning quality of the formally developed grand-canonical extension to the \textsc{rbm},
we shall not further discuss sampling options and evaluate the loss functions on the mean-field theoretic ansatz.
 

%

\paragraph{Computational simplications.}
In developing the theoretical framework for the \vrbm we have been appealing to the large order of  hidden units, $K\gg N$, to render certain designer choices in the form of the chemical potential equivalent.
The regime of large $K$ is a key feature to ensure the desired elimination of the length of the hidden layer as a hyperparameter. 
One might wonder whether  such a regime is in practice feasible, beyond a mere  theoretical playground.
On the one hand, it turns out that $K$ does not need to be that large for the \vrbm to work in a satisfactory self-regularizing manner. At least for the systems we have considered, finite-$K$ effects appear way beyond $\varepsilon_\text{test}$, for very reasonable values of $K$, not influencing thus the quality of learning.
On the other hand,  larger networks become quickly --\,already from the first \textsc{cd} steps\,-- exponentially suppressed in $p(z\vert\vset)$,
thus setting 
hidden expectation values $\VEV{h^a}$ for larger $a$ effectively to zero (see Eq.~\eqref{h_VEV}).
Hence, one needs to calculate the \textsc{cd} algorithm~\eqref{a_CDupdate} to update the \vrbm parameters only for $K_\text{eff}\ll K$ hidden units. This computational simplification considerably speeds up the training while allowing us to take the maximum number of hidden units even larger and verify the  independence of the learning procedure from $K$ (to the desired level of accuracy).
%
 

\subsection{The Ising model}
\label{ssc:Ising}

First, we choose to train our \vrbm on a system from 
statistical physics where the target distribution $q(\vset)$ is extracted by sampling spin configurations on a lattice at certain temperatures $T$. Depending on the number of space-time dimensions and the amount of relevant symmetries the physical system can be under a (partial at least) analytic control.
%
In the physics literature, a paradigm statistical system is the Ising model with nearest neighbour interactions.
The first non-trivial behaviour of the Ising theory that exhibits a phase transition at a critical temperature $T=T_c$ from a ferromagnetic to a paramagnetic phase in infinite volume  manifests~\cite{onsager1944crystal} in two space-time dimensions.  
In absence of external magnetic fields, the partition function of this system up to two space-time dimensions can be calculated exactly~\cite{baxter2016exactly}.

\begin{figure}[t]
\begin{center}
\includegraphics[width=16cm,height=16cm,keepaspectratio]{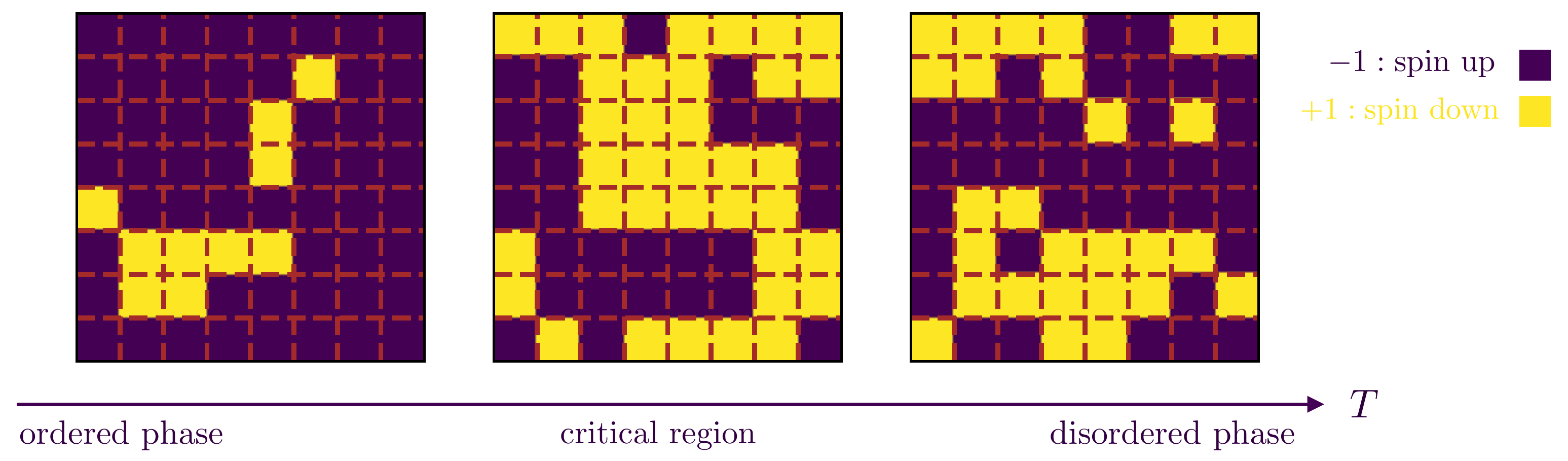}
\caption{Ising configurations  on an $8\times8$ lattice sampled via \textsc{mcmc} simulations. Spins at neighbouring sites  try to align along the same direction. 
The order parameter, the net magnetization, which is given as the sum of all $64$ spins is non-vanishing in the ordered phase below the critical temperature and (abruptly in infinite volume) vanishes when the system enters the disordered phase at higher temperatures.}\label{fig:IsingCongigs}
\end{center}
\end{figure}

For concreteness, we take the Ising theory to live on a square lattice of length $L$ described by a spin matrix
\equ{
s^{\ga\gb}\in \{-1,1\} \quads{\text{with}} \ga,\gb=1,...,L 
}
with periodic boundary conditions $s^{L+1,\gb}\equiv s^{1,\gb}$ and 
$s^{\ga,L+1}\equiv s^{\ga,1}$. The nearest-neighbour Hamiltonian reads
\equ{
\label{IsingHamiltonian}
H_\text{Ising} = - J\sum_{\langle (\ga\gb) ,(\gamma\delta)\rangle } s^{\ga\gb} s^{\gamma\delta}
}
where the sum at each lattice point is taken over the four --\,in case of square lattices\,-- directly neighbouring  sites. 
$J$ is the Ising coupling whose sign dictates the (anti)ferromagnetic structure of spin configurations. 
In the thermodynamic limit $L\rightarrow\infty$, boundary effects become negligible and universality ensures the manifestation of the transition from the ordered to a disordered phase independently of the microscopic details.
In practice, we observe most important features (like a peak in heat capacity at $T\approx T_c$) signalling the aforementioned phase transition already for $L=8$. 
The aim of this section is to train our \vrbm on this statistical system to test whether our self-regularizing \textsc{ml} algorithm can distinguish the physical interactions (leading to cluster formations as depicted in Figure~\ref{fig:IsingCongigs})  from mere thermodynamic fluctuations (cf.\ noisy high-temperature data in Figure~\ref{fig:IsingCongigs}).

To extract the relevant physics of the Ising system the \vrbm should be exposed to spin configurations $s$ sampled at various temperatures  below and above the critical region.  
As noticed in~\cite{Iso:2018yqu} (a behaviour also verified for our \textsc{vrbm}), even without training an \rbm in the vicinity of $T_c$, but only below and above,   the generative model is still able to capture the physics signalling a phase transition.
Via Markov chain Monte Carlo (\textsc{mcmc}) simulations we thus produce a large number of spin configurations $s$ at various temperatures $T=0.1,0.2,..,4.5$ which we split\footnote{This way of splitting is to facilitate comparison with the handwritten dataset used for training the \vrbm in the following paragraph. In fact, already $10\,000$ training samples suffice so that the \vrbm extracts the physics of the target system at a satisfactory level.} into a training and test set of $60\,000$ and $10\,000$ samples, respectively.
Our target distribution $q(\vset)$ thus extends over the simulated Ising domain with $v^{(\ga-1)\cdot L+\gb}=s^{\ga\gb}$.

At this stage, revealing essentially the final outcome is in order.
In~\cite{Iso:2018yqu,2017arXiv170804622M}
the learning capacity of the ordinary \rbm on the Ising model has been extensively studied showing that there exist three distinct learning phases depending on the relation between the number of hidden units $z_0$ and the total number of spins $N=L\times L$. 
For $z_0<N$ the \rbm does not have enough resources to fully learn the target distribution from the Ising system. 
{In this regime with less hidden than observable units, an appropriately trained generative model has still learned important features of the underlying Ising theory. In particular, the \rbm flow seems to trigger a flow of spins very reminiscent of the Renormalization Group~\cite{Funai:2018esm}.
We plan to come back to this tantalizing finding in conjunction with varying number of hidden units and different training metrics in a later work.}
When $z_0=N$ the \rbm fully learns to reproduce the Ising theory of nearest-neighbour interactions at any temperature.
Finally, over-learning starts to occur for $z_0>N$ and the \rbm increasingly learns irrelevant noice of thermodynamic fluctuations with increasing number of hidden units.

\paragraph{Specifics of the training.}
For our purposes, we have trained various generative models at a chemical potential $\mu$ of slightly different designs which fall within the class described in Section~\ref{ssc:ChemicalPotential}.
Since such designer choices do not influence the final outcome for sufficiently large number $K$ of available hidden units, we concentrate here, for concreteness and clarity of presentation, on the ansatz~\eqref{ChemicalPotential_FunctinalForm} with $p=1$.
Our mean-field theoretic ansatz~\eqref{kcd_step} to train our model by numerically solving the extremization problem posed in paragraph~\ref{ssc:CD} converges to a satisfactory level already after $k=2$.
Once exposed to various temperatures, any self-regularizing generative model should be able to track down the clearly defined regime of optimal training, $z=N$, detecting the physical interactions of nearest-neighbouring spins.  

\begin{figure}[t]
\begin{center}
\includegraphics[width=17cm,height=17cm,keepaspectratio]{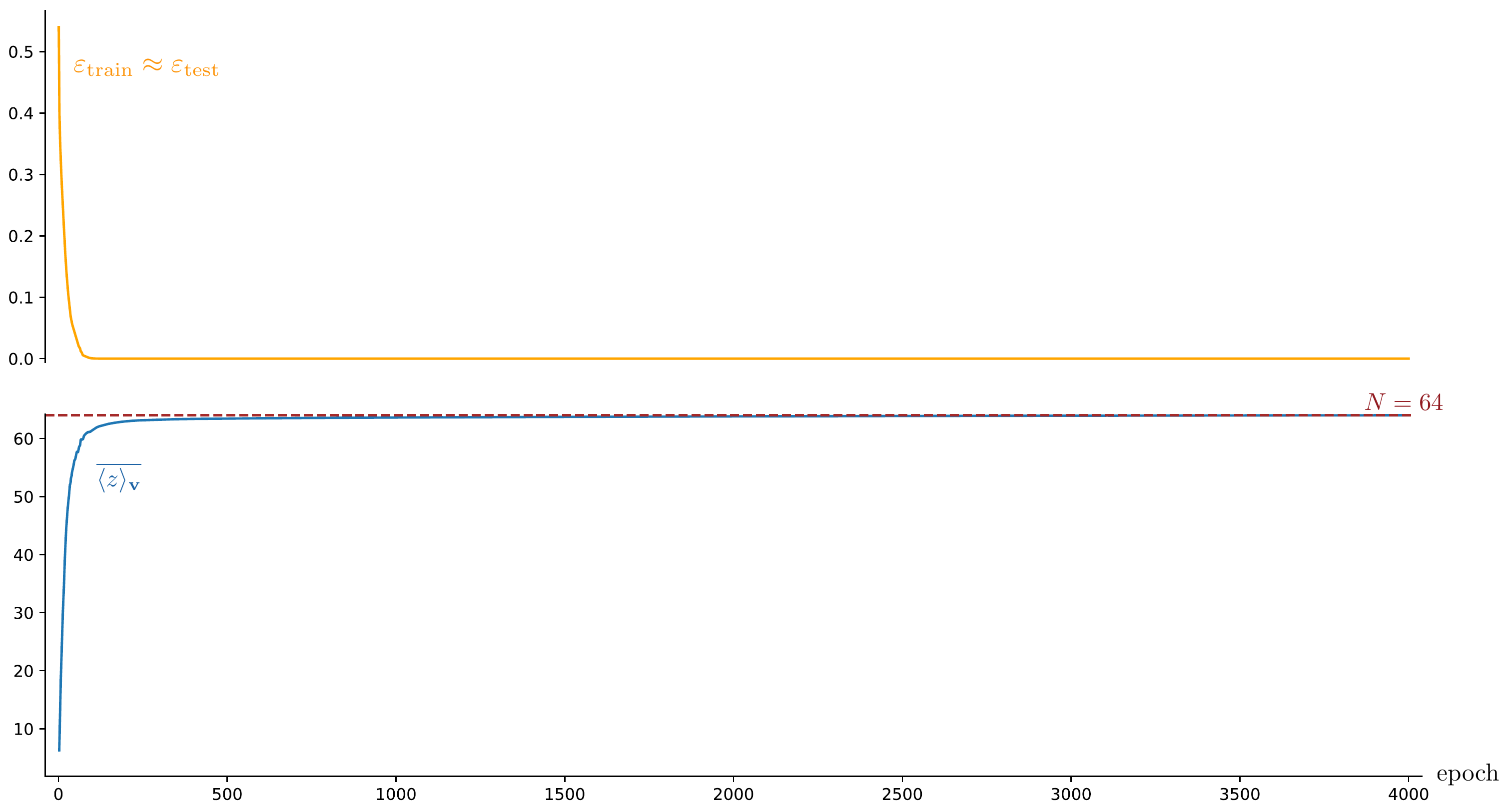}
\caption{The train~\eqref{TrainErrorDEF} and test~\eqref{TestErrorDEF} error  as well as the average of the expectation value of the number of hidden units~\eqref{zAverage}  are plotted while training the \vrbm with $K=2\,000$ for $4\,000$ epochs.}\label{fig:Ising:LossVSz}
\end{center}
\end{figure}

Indeed,
as seen in the upper Plot~\ref{fig:Ising:LossVSz}, our \vrbm with  $K=2\,000$  hidden units appropriately learns after a couple of epochs the two-dimensional Ising system. The exceedingly small training and test quadratic error, $\varepsilon_\text{train}=\varepsilon_\text{test}=\order{10^{-13}}$ as calculated without rounding by Eq.~\eqref{TrainErrorDEF} and~\eqref{TestErrorDEF}, respectively,  effectively coincide throughout most of the training.
To comprehend this order of magnitude,
note that an ordinary \rbm with (fixed) number of hidden units already of order $z_0=\order{100}$
leads for the same Ising configurations of total lattice size $N=64$ to a considerable test error $\varepsilon_\text{test}=\order{0.1}$.
In fact, one does not even need to take $K$ that large to observe this self-regularizing character. A \vrbm with $K=200$ only, effectively demonstrates the same learning curve as in Figure~\ref{fig:Ising:LossVSz} once trained over the same data set.
Sampling from those trained \vrbm leads to perfect reconstructions of the original Ising data.
The chemical potential Eq.~\eqref{ChemicalPotential:ConceptualAnsatz} which is dynamically determined throughout training remains an order one quantity; specifically for the Ising model we get $\mu \approx 1.01$.

\begin{figure}[t]
\begin{center}
\includegraphics[width=12cm,height=12cm,keepaspectratio]{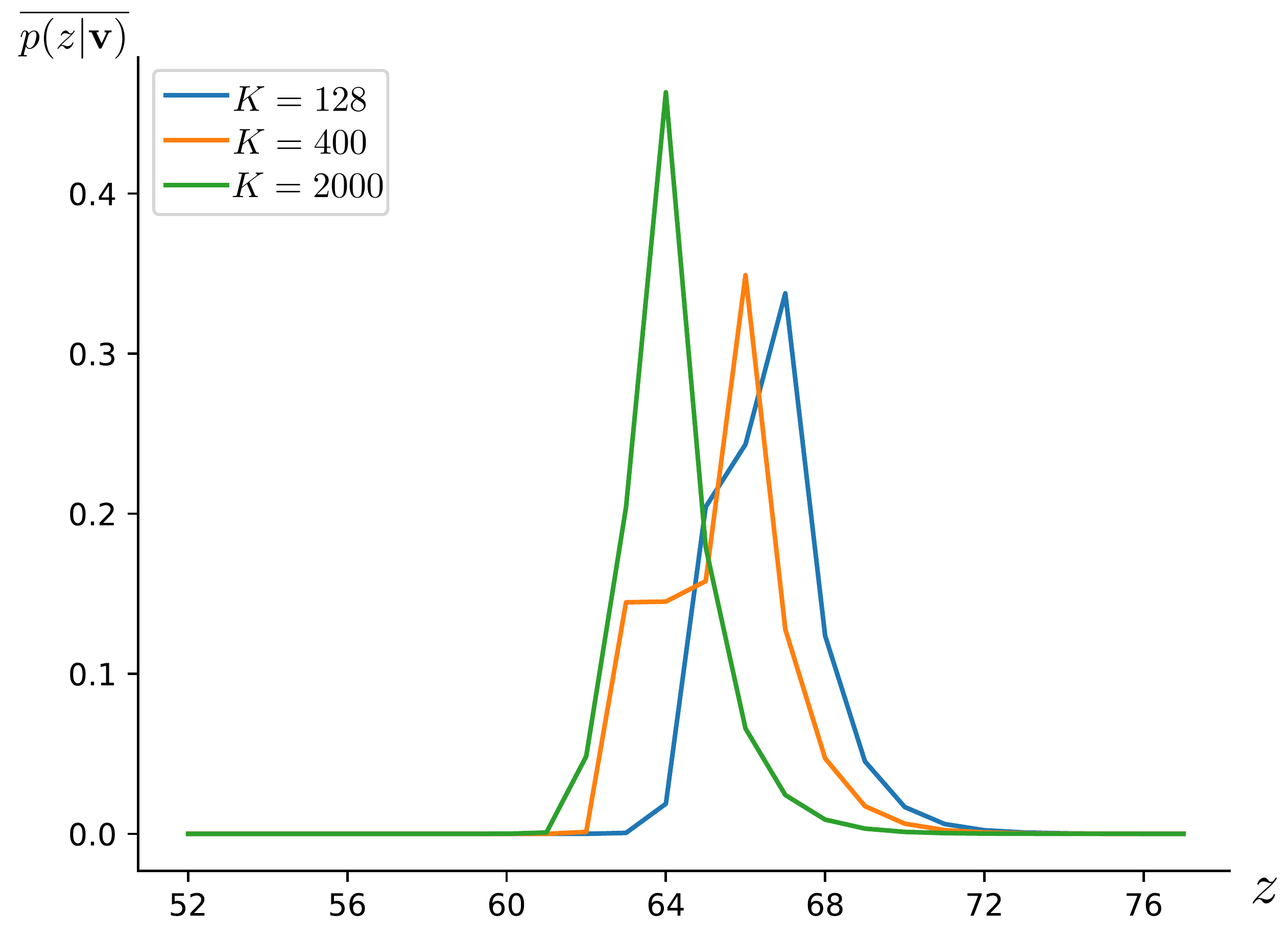}
\caption{
The average of the conditional probability Eq.~\eqref{average_z_ConditionalProbability} over the Ising data from the square lattice of length $L=8$ is plotted for different lengths $z$ of the hidden layer in the model with maximally $K=128,400,2\,000$ hidden units.
The probability clearly exhibits a peak around $N=L^2=64$. For clarity, the plot concentrates on the region around $z=N$ given that networks of sizes away from it quickly get exponentially suppressed.}\label{fig:Ising:z_ConditionalProbability}
\end{center}
\end{figure}

To understand what the \vrbm has actually learned and how it did that, we plot in the lower part of Figure~\ref{fig:Ising:LossVSz} the average of the expectation for the number of hidden units (see Eq.~\eqref{z_conditionalVEV}) conditioned on the Isisng data,
\equ{
\label{zAverage}
\overline{\VEV{z}_\vset} = \sum_{\vset} q(\vset) \VEV{z}_\vset 
~,
}
over the course of learning epochs, together with the learning error. We clearly see that the model starts learning when $\VEV{z}_\vset$ becomes $\order{N}$.
Along similar lines, Figure~\ref{fig:Ising:z_ConditionalProbability} depicts the conditional probability $p(z\vert\vset)$ defined in Eq.~\eqref{z_conditionalProbability}, averaged over the dataset, 
\equ{
\label{average_z_ConditionalProbability}
\overline{p(z\vert\vset)} = \sum_\vset q\left(\vset\right) p\left(z\vert\vset\right)
}
for different number of available resources, $K=128,400,2\,000$, ranging from $K=2N$ towards the formally desired regime $K\gg N$.
On the one hand, it is clear that 
networks of hidden size between $z\approx60$ and $z\approx70$ receive a significant probability to participate into feature extraction.
Thus, the \vrbm still slightly over-learns due to finite\,--\,$K$ effects, with a test error $\varepsilon_\text{test}$ though, which is essentially zero for any practical purpose.
On the other hand, we observe that the curve of $p(z\vert\vset)$  becomes narrower around the critical value  $z=64$ with increasing $K$.
This is nothing but a manifestation of ``the law of large numbers'': at the theoretical limit $K\rightarrow\infty $ we anticipate the \vrbm to precisely pick $z=64$ as the most optimal size of the hidden layer to learn the provided Ising configurations.

\begin{figure}[t]
\begin{center}
\includegraphics[width=17cm,height=17cm,keepaspectratio]{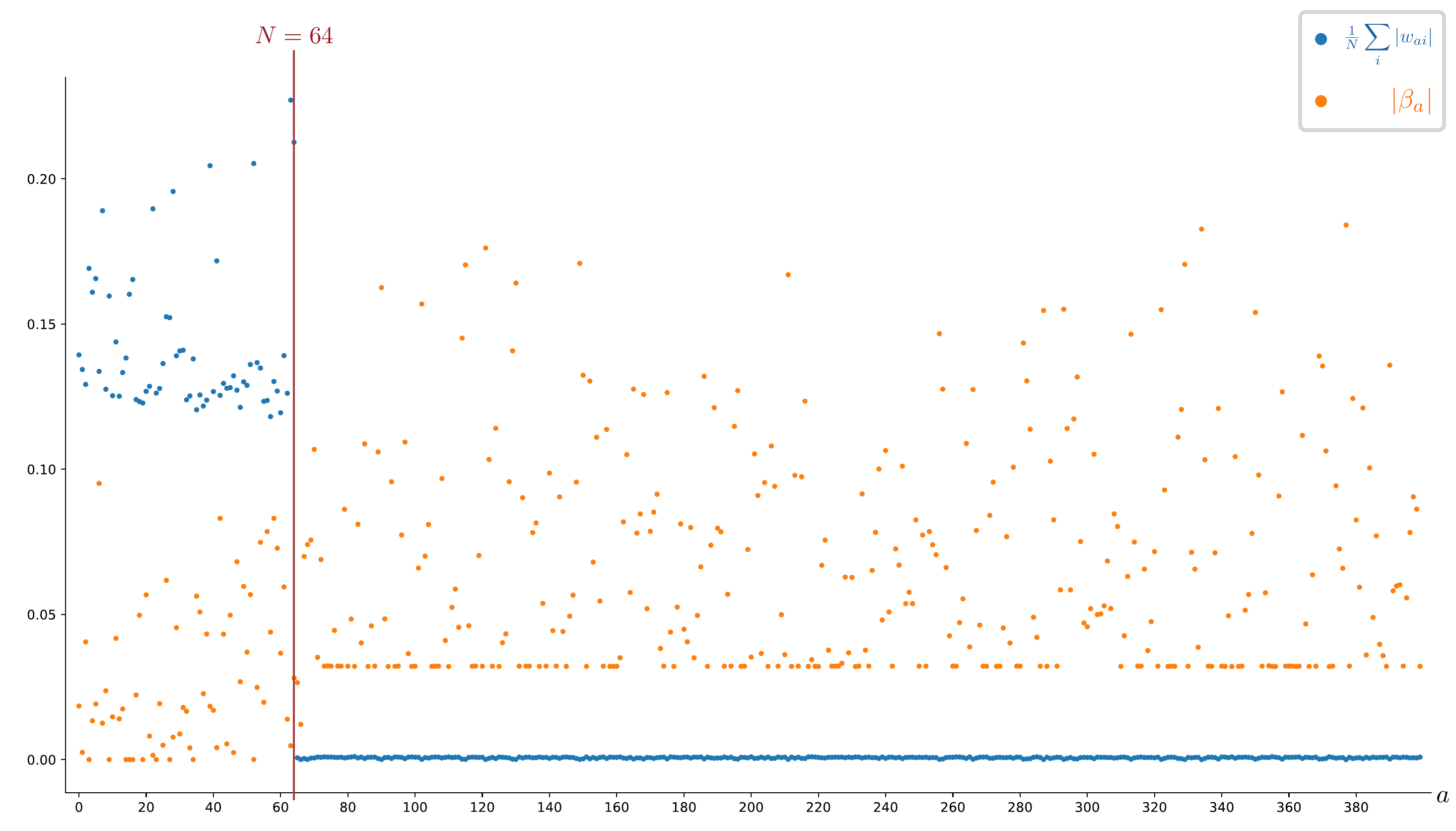}
\caption{The scatter diagram depicts the row-average over the absolute value of weights $w_{ai}$ defined in  Eq.~\eqref{AbsoluteWeightAverage} together with the absolute value of hidden biases $\beta_a$ for the first $a=1,...,400$ after training a \vrbm with $K=2\,000$ for $4\,000$ epochs.}\label{fig:Ising:weights_VS_biases}
\end{center}
\end{figure}

\begin{figure}[t]
\begin{center}
\includegraphics[width=15cm,height=15cm,keepaspectratio]{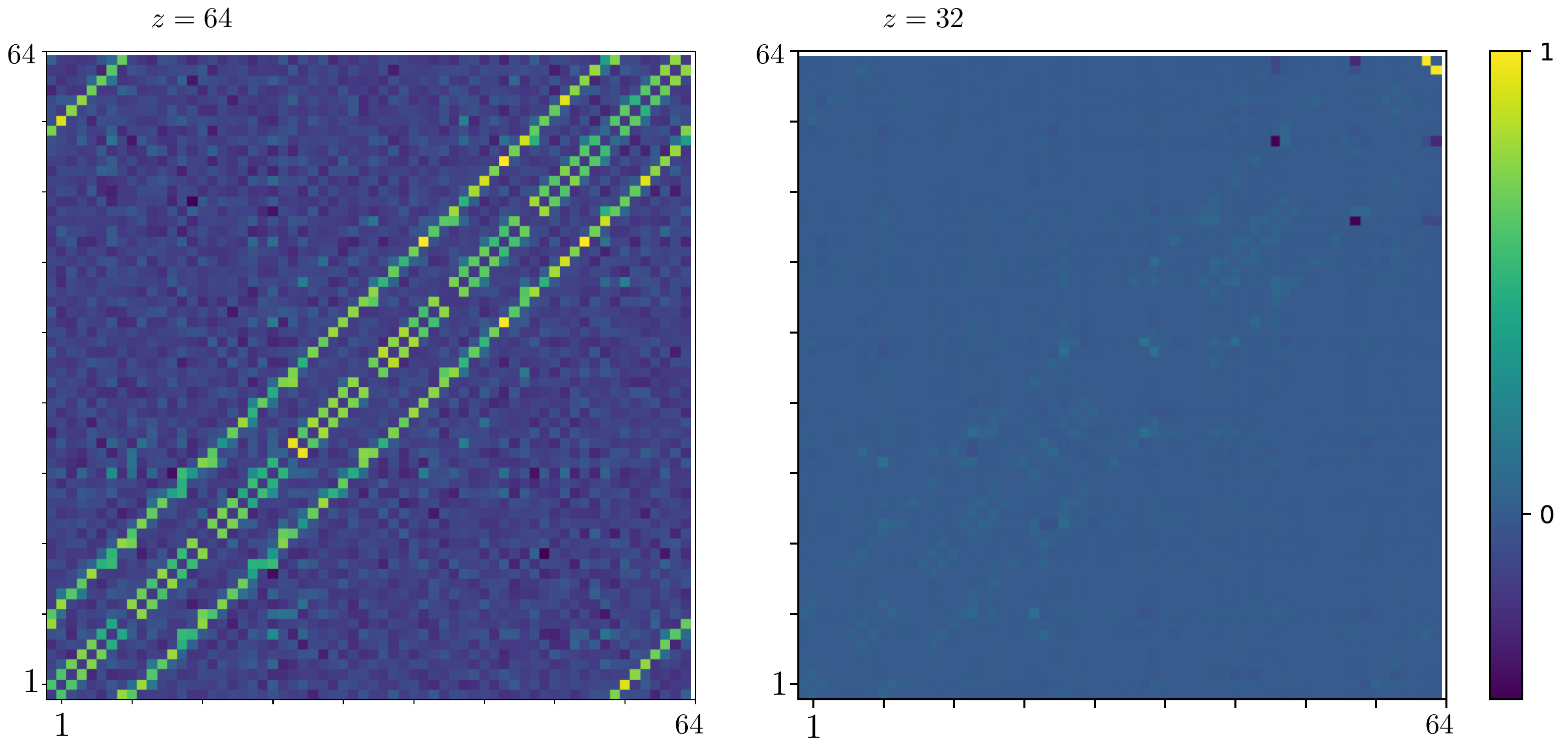}
\caption{The heat map chart represents the interaction matrix $S^z$ defined in Eq.~\eqref{Ising:interactionMatrix}, normalized over its largest absolute value, for two networks with hidden sizes $z=32,64$ using the  parameters of the \vrbm with $K=2\,000$ trained over $4\,000$ epochs on the Ising data from the $64\times64$ lattice.}\label{fig:interactionMatrix}
\end{center}
\end{figure}

To further comprehend the behaviour of the grand-canonical generalization of \rbm under training, 
we examine the value of the trained parameters from the perspective of the hidden units. 
The meaningful quantities to look at are the hidden biases $\bias_a$ and the average
\equ{
\label{AbsoluteWeightAverage}
\overline{\abs{w_a}} = \sfrac1N\sum_{i=1}^N \abs{w_{ai}}
}
entering the chemical potential via Eq.~\eqref{FundamentalEnergyQuantum}.
For the first $a=1,...,400$ hidden units of the trained \vrbm with $K=2\,000$ these are plotted in Figure~\ref{fig:Ising:weights_VS_biases}.
As theoretically anticipated, we recognize that the \vrbm has effectively set to zero all weights $w_{ai}$ for $a>N$ in accordance with its self-regularizing character.
In the language of paragraph~\ref{ssc:ChemicalPotential} thus, $K_\text{eff}\approx N$. The latter $1\,600$ hidden units not depicted in Figure~\ref{fig:Ising:weights_VS_biases} follow an evident pattern for $a>K_\text{eff}$ and decouple from the feature detector (see schematic depiction in Figure~\ref{fig:vRBM_schema}).
Incidentally, due to the aforementioned regularizing character also at smaller $K$, the depicted profile in Figure~\ref{fig:Ising:weights_VS_biases} looks effectively the same also when $K=400$ and even $K=128$. 
Most interestingly, the value of the first $K_\text{eff}$ hidden biases $\bias_a$ is smaller than the corresponding weight scale set by Eq.~\eqref{AbsoluteWeightAverage}. 
Thus, these $\bias_a$ do not participate much in the modelling of  Ising interactions performed by the first $K_\text{eff}\times N$ weights $w_{ai}$ (in the sense of Figure~\ref{fig:interactionMatrix}).  
The model then uses the remnant $(K-K_\text{eff})$ biases to adjust the value of the chemical potential $\mu\equiv\mu(w_{ai},\bias_a)$ and  its regularizing effect without crucially interfering with the actual feature detection.

Another quantity of interest that is meaningful to examine  for energy-based generative models is the free energy $F(\vset)$ defined  in Eq.~\eqref{GrandCanonical:FreeEnergy}.
An estimation of its value is given by the grand-canonical expectation value of the free energy introduced in Eq.~\eqref{GrandCanonical:FreeEnergy1a},
\equ{
\label{AverageFreeEnergy}
\VEV{F(\vset,z)} = \sum_{z=1}^K p\left(z\vert\vset\right) F(\vset,z) 
~,
}
under the conditional probability Eq.~\eqref{z_conditionalProbability} of our model. 
For binary systems it is straight-forward~\cite{2018arXiv180308823M}
to expand $F(\vset,z)$ in powers of $v^i$ and resum using that $(v^i)^2=1$.
In~\cite{Cossu:2018pxj}
the leading  terms
in the spin  expansion were computed, 
\equ{
\label{FreeEnergy:Resum}
F(\vset,z) = -\sum_{i=1}^N J^z_i v^i
-\sum_{i,j}^N v^i\,S_{ij}^z\,v^j
+...~, 
}
in terms of the spin current $J^z$ and the correlation matrix $S^z$  of a network with $z$ hidden units.
For reasonably small values of the trained parameters $\gl$
the two scale as 
%
\equ{
\label{Ising:interactionMatrix}
J^z_i = c_i + \sum_{a=1}^z b^a w_{ai} + \order{\gl^3}
\qand
S_{ij}^z =
\sum_{a=1}^z w_{ai}\, w_{aj}
+ \order{\gl^3}
\quads{\text{for}} i\neq j
}
and $S_{ii}=0$, since $(v^i)^2=1$ gives an irrelevant constant energy shift.

Already in this crude approximation the current $J^z$ evaluated on the trained parameters $\gl$ appears subleading to the spin-spin interaction term (cf.\ also Figure~\ref{fig:Ising:weights_VS_biases}), as  anticipated given that the Ising data was originally sampled at zero external magnetic field. 
In Figure~\ref{fig:interactionMatrix}, we  draw the correlation matrix $S^z$ for different hidden-layer sizes $z$.
(Note that the heat map chart of $S^z$ for $z>N$ will not look different from $z=64$, as $w_{ai}\approx0$ for $a>N$.)
After our preceding discussion, there is no surprise that  networks with $z<N$ cannot satisfactory learn the input data. Their contribution to the approximation~\eqref{AverageFreeEnergy} to grand-canonical free energy is exponentially suppressed by $p(z\vert\vset)$ as seen in Figure~\ref{fig:Ising:z_ConditionalProbability}.
In contrast, approaching $z=N$ the nearest-neighbour  structure of the Ising data becomes apparent, at least to quadratic order in the spin expansion.
%
The latter are precisely the networks which get significantly  selected  by Eq.~\eqref{AverageFreeEnergy} to participate in forming our estimation of the free energy of the \vrbm model.
In a future work, we plan to come back to the intriguing question of the order-by-order equivalence among the energies from the trained \vrbm and the Ising model by formal resummation of the appropriate free energy Eq.~\eqref{GrandCanonical:FreeEnergy} in the spirit of Eq.~\eqref{FreeEnergy:Resum}. 

\subsection{The dataset of handwritten digits}
\label{ssc:MNIST}

\begin{figure}[t]
\begin{center}
\includegraphics[width=15cm,height=15cm,keepaspectratio]{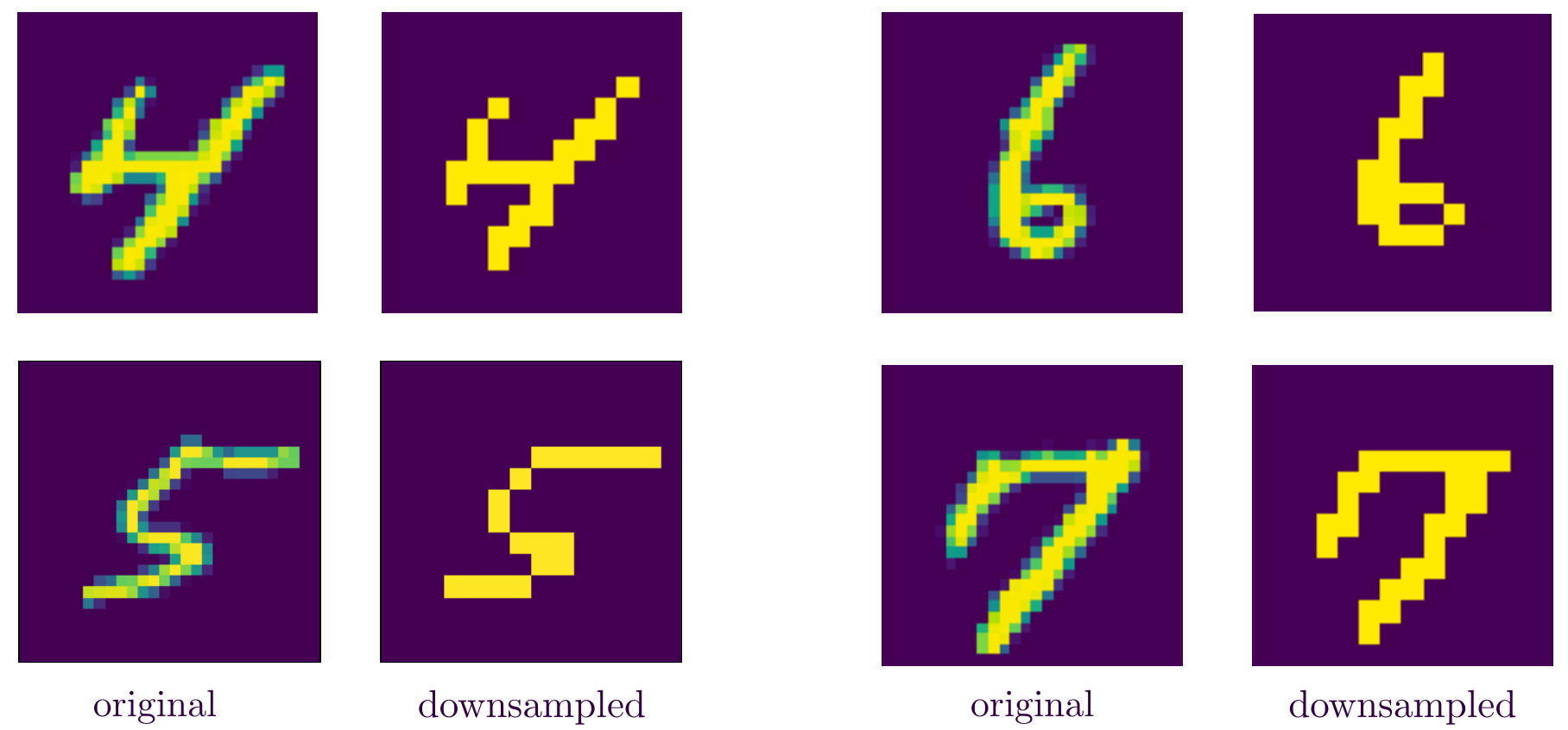}
\caption{Examples from the \textsc{mnist} data set downsampled to $14\times14$ pixel size and binarized.}\label{fig:MNIST_digits}
\end{center}
\end{figure}

As a first step in benchmarking \textsc{ai} models it is customary to draw samples from the \textsc{nist} database of ``Handprinted Forms and Characters''.
By now, a typical list of examples used throughout \textsc{ml} literature is the \textsc{mnist} data set~\cite{726791} of handwritten digits from 0 to 9.
It consists of $60\,000$ training and $10\,000$ test  preprocessed images given in \textsc{rgb} format with an original $28\times28$ pixel resolution.
Let such a sample image be described by an integer matrix of  \textsc{rgb} pixel intensities $S^{AB}\in[0,255]$ with $A,B=1,...,28$.

For our demonstrative purposes, it suffices to consider a reduced version  of the \textsc{mnist} data by downsampling to a $14\times14$ image version. 
This can be done via so called \textit{max} or \textit{mean pooling}~\cite{masci2011stacked,sohn2012learning}.
Concretely, each $2\times2$ block of pixels in the original pixel matrix $S$ is replaced by either their average or their maximum leading for example to
\equ{
\sigma^{\ga\gb} := \text{max}\{ S^{2\ga-1,2\gb-1},S^{2\ga,2\gb-1},S^{2\ga-1,2\gb},S^{2\ga,2\gb} \}
\quads{\text{with}} \ga,\gb=1,...,L
~,
}
and similarly for their average.
In our case, $L=14$.
To smoothen the resulting image $\gs$ and to reduce its size
while preserving important information for the feature detector it is 
possible to apply additional space-convolution filters (adding a padding frame where necessary) like 
\equ{
\tilde \gs^{\ga\gb} :=  \sfrac14\left(\gs^{\ga\gb}+\gs^{\ga,\gb+1}+\gs^{\ga+1,\gb}+\gs^{\ga+1,\gb+1}\right)~,
}
which captures an average pixel intensity in overlapping $2\times2$ blocks. 
Throughout the down-sampling and convolutional process we always make sure that the pixel center of mass, defined by 
\equ{
\begin{pmatrix}
	\ga_\text{cm}\\
	\gb_\text{cm}
\end{pmatrix}
= 
\left(\sum_{\ga',\gb'=1}^L \gs^{\ga'\gb'}\right)^{-1}
\sum_{\ga,\gb=1}^L \gs^{\ga\gb} 
\begin{pmatrix}
	\ga\\
	\gb
\end{pmatrix}
~,
}
coincides with the geometrical center of the image located 
at $\left(L/2,L/2\right)$ in order to filter out translational symmetry.
Furthermore, to make contact with the preceding paragraph on the Ising model we turn to black and white images via binarizing all pixels simply by rounding each normalized pixel 
to its nearest integer (i.e.\ 0 or 1). Hence, our training input is given by $v^{(\ga-1)\cdot L+\gb}=s^{\ga\gb}$ with
\equ{
s^{\ga\gb} := 2\round{\frac{\gs^{\ga\gb}}{255}}-1~,
}
where we again arrange for $s^{\ga\gb}\in\{-1,1\}$ consistent with the conventions introduced in Section~\ref{ssc:vRBM}.
%
An example of the target images we are going to use in the following paragraph to train our \vrbm is given in Figure~\ref{fig:MNIST_digits}.

\paragraph{Specifics of the training.}
\begin{figure}[t]
\begin{center}
\hspace{-0.3cm}
\includegraphics[width=17cm,height=17cm,keepaspectratio]{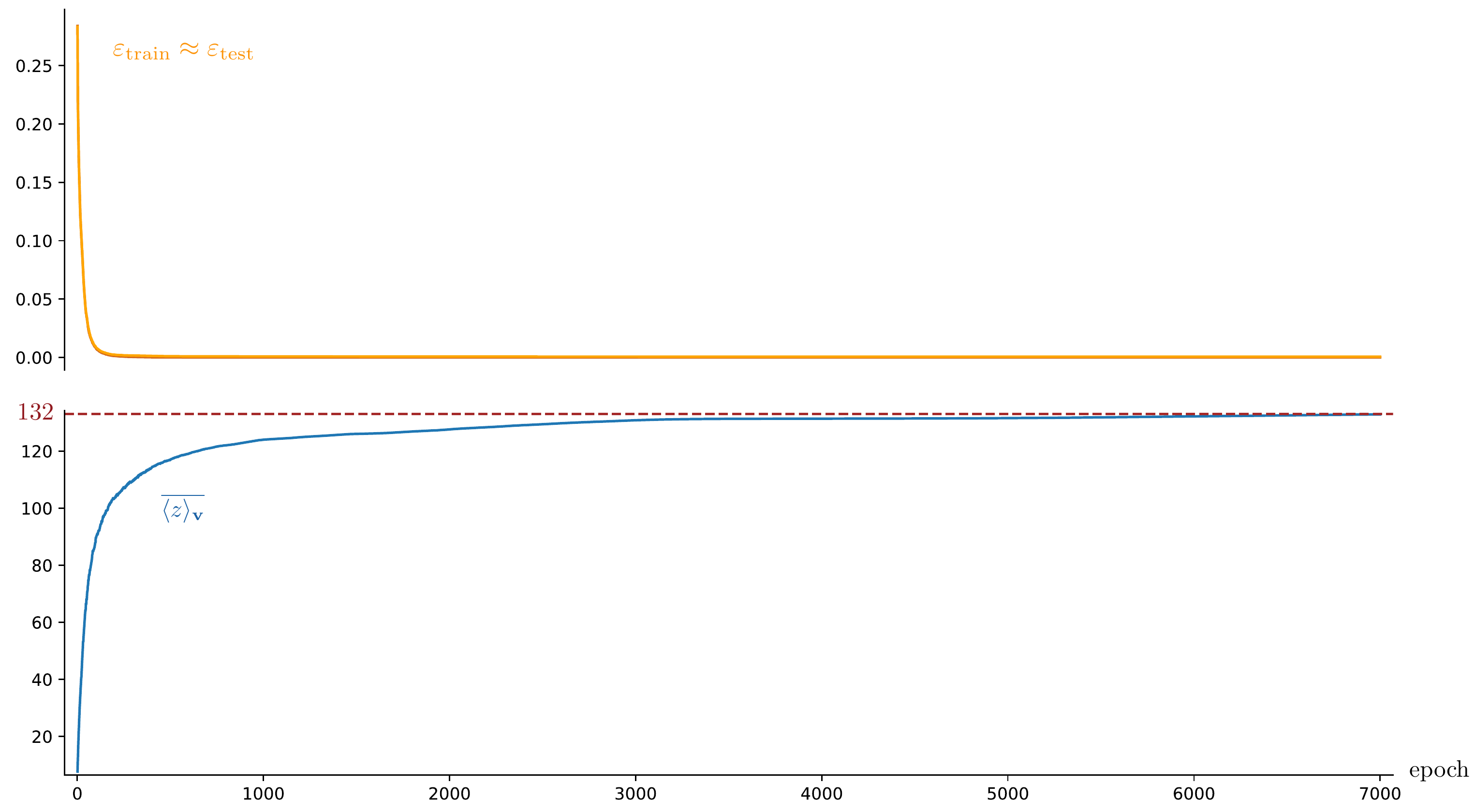}
\caption{The train~\eqref{TrainErrorDEF} and test~\eqref{TestErrorDEF} error  as well as the average of the expectation value of the number of hidden units~\eqref{zAverage}  are plotted while training the \vrbm with $K=900$ for $7\,000$ epochs.}\label{fig:MNIST:LossVSz}
\end{center}
\end{figure}
Similar to the Ising model, we train \vrbm of different designs ($p=1,2$ in the chemical potential Eq.~\eqref{ChemicalPotential:ConceptualAnsatz}) on the downsampled \textsc{mnist} data of observable size $N=L^2=196$
and recover the law of large numbers.
With increasing number of available hidden resources the outcome stabilizes and effectively becomes independent of $K$.  
The train and test loss errors, Eq.~\eqref{TrainErrorDEF} and~\eqref{TestErrorDEF}, quickly become $\order{10^{-8}}$ and $\order{10^{-5}}$, respectively, signalling a very good convergence and a small over-learning. As a reference, an ordinary \rbm with $z=200$  trained on the exact same data without any form of regularization has a test error of $\order{10^{-3}}$.  
In Figure~\ref{fig:MNIST:LossVSz}, the loss errors are plotted over the learning epochs as well as the average of the expectation value for the size of hidden layer Eq.~\eqref{zAverage}. Again, our \textsc{ml} model starts learning when networks of a given size around $\overline{\VEV{z}_\vset} \approx 132$ become more and more favourable. 
%
The chemical potential is dynamically determined during training to $\mu\approx 1.1$.

\begin{figure}[t]
\begin{center}
\includegraphics[width=17cm,height=17cm,keepaspectratio]{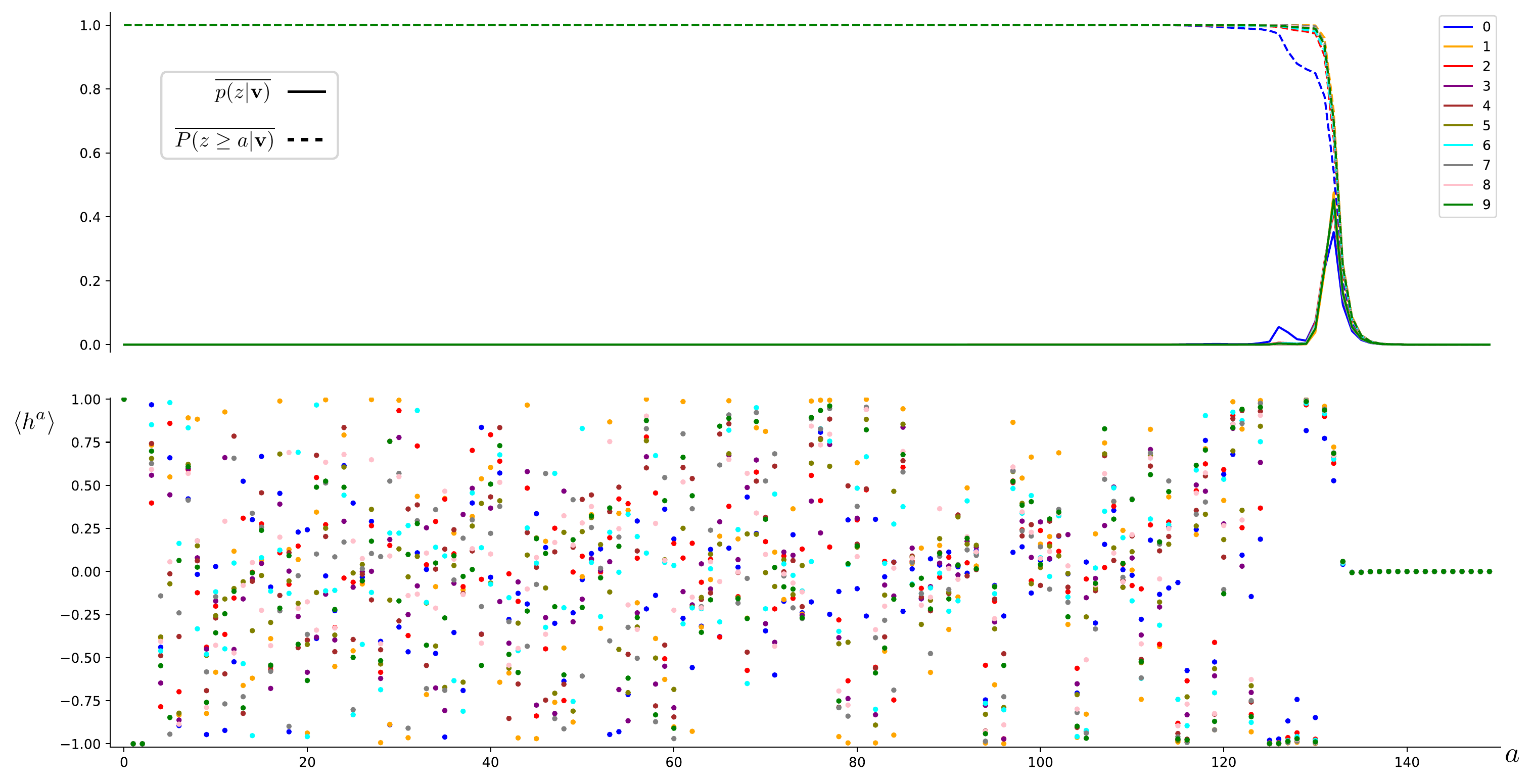}
\caption{Continuous lines in the upper plot depict the average of conditional probability Eq.~\eqref{z_conditionalProbability} over the \textsc{mnist} data for each handwritten digit 0,...,9 (the color pattern applies also to the scatter plot) of a trained \vrbm with $K=900$ hidden units. Networks with bigger $z$ are exponentially suppressed and are not presented in the graph for clarity, $z,a=1,...,150$.  
Dashed lines depict the corresponding ccdf in Eq.~\eqref{ccdf}. 
The lower scatter diagram presents the expectation value of hidden units as deduced from Eq.~\eqref{h_VEV} digit-wise conditioned on the input data.}\label{ccdf_zProbability_VS_hEval}
\end{center}
\end{figure}

To obtain a better feeling of what the \vrbm has learned from the dataset of handwritten digits it is sensible to look at the average of conditional probability Eq~\eqref{average_z_ConditionalProbability} over each digit from \textsc{mnist}, separately.
From the upper part of Figure~\ref{ccdf_zProbability_VS_hEval} it becomes clear that for all digits the most probable size of the hidden layer coincides with the (rounded) average
$\round{\overline{\VEV{z}_\vset}} = 132$.
Still, for an ``easy'' digit like  zero there appears a lower pump already before $z$ approaches the region of $\overline{\VEV{z}_\vset}$. 
Such a profile for ${p(z\vert\vset)}$ is typical for a system coming from everyday life and is comparatively richer to the Ising paradigm of the previous section, where all data came from the same microscopic Hamiltonian Eq.~\eqref{IsingHamiltonian}.

To avoid any confusion, the scatter diagram in the lower part of Figure~\ref{ccdf_zProbability_VS_hEval} emphasizes that the probability $p(z\vert\vset)$ according to which the size of the hidden layer gets selected is a different concept from the actual feature detection happening at the level of $\VEV{h^a}_\vset$. Depending on the digit they track, hidden units $h^a$ are more or less likely to get activated with a given sign.
Of course, the two concepts are intimately connected via Eq.~\eqref{h_VEV}: the complementary cumulative distribution associated to $z$ modulates the profile of $\VEV{h^a}_\vset$.
The hidden units $h^b$ with $b>K_\text{eff}$ of larger hidden networks, whose probability to be selected becomes exponentially suppressed, remain deactivated.
Stochastically, the equivalent statement would be that those $h^b$ receive an equal probability to be $\pm 1$, behaving like free units, as they have decoupled from connected Boltzmann machine in Figure~\ref{fig:vRBM_schema}.

\begin{figure}[t]
\begin{center}
\hspace{-0.3cm}
\includegraphics[width=17cm,height=17cm,keepaspectratio]{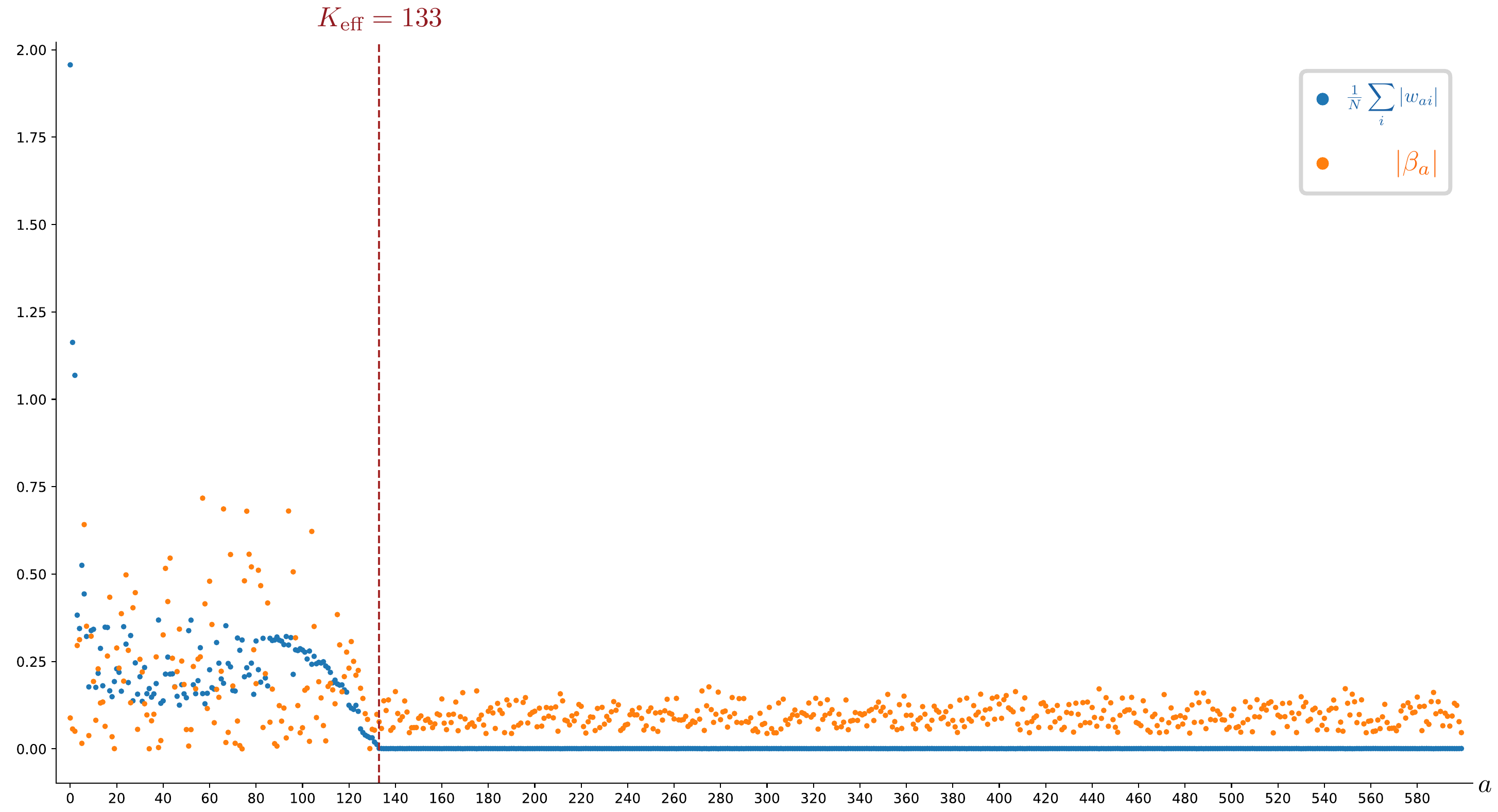}
\caption{The scatter diagram depicts the row-average over the absolute value of weights $w_{ai}$ defined in  Eq.~\eqref{AbsoluteWeightAverage} together with the absolute value of hidden biases $\beta_a$ after training a \vrbm with $K=900$ for $7\,000$ epochs.}\label{MNIST:weights_VS_biases}
\end{center}
\end{figure}

Indeed, the weights for $a> K_\text{eff}=133$ have been effectively regularized to zero, as deduced from Figure~\ref{MNIST:weights_VS_biases}.
Hence, the corresponding hidden units to the far right of the plot decouple in the sense of the schematic depiction in Figure~\ref{fig:vRBM_schema}.
The scatter plot~\ref{MNIST:weights_VS_biases} follows the same regularizing concept as the Ising plot~\ref{fig:Ising:weights_VS_biases} and we refer the reader to the corresponding paragraph in Section~\ref{ssc:Ising} about the Ising model.
An obvious difference observed in the two plots for the first $a=1,...,K_\text{eff}$ hidden units lies in the different character of the systems:
Concerning the \textsc{mnist} dataset, hidden biases $\bias_a$ actively participate in feature detection in contrast to the  Ising scenario in absence of  external magnetic fields.

\section{Conclusions and outlook}

In this paper, we have considered (shallow) restricted Boltzmann machines at a finite chemical potential $\mu$. In principle, such a grand-canonical extension of the \rbm performs feature extraction from a target distribution $q(\vset)$  by invoking hidden layers $\{h^a\}_{a=1,...,z}$ of various length $z=1,...,K$ to model interactions among $N$ observables units $v^i$, where in principle $K\gg N$.
We have concentrated on an intuitive choice of the chemical potential as a function of a ``vacuum'' energy, which essentially measures the added norms of the weight matrix $\weight_{ai}$ and bias vector $\bias_a$ per unit $h^a$ actively participating in feature extraction.
The appropriately trained \vrbm at such chemical potential $\mu\equiv\mu(\abs{\weight_{ai}},\abs{\bias_a})$ is able to track down (up to $N/K$--suppressed effects) the  optimal length of the hidden layer to model provided data points.
To do so, the \vrbm mainly uses the biases $\bias_c$ of disconnected ($\weight_{ci}=0$)  hidden units $h^c$ to regulate the number of hidden units $K_\text{eff}$ which actively participate in feature extraction, manifesting the self-regularizing character of the model.

Incorporating this regularizing character in the form of a chemical potential has many advantages, besides the formal simplicity of Legendre-transforming to the grand-canonical ensemble, which allows us to keep most of the techniques implemented to train ordinary Boltzmann machines intact.
By maximizing  the expectation value of $\log$-probability of the modelled data under $q(\vset)$ 
the value of $\mu$ is dynamically fixed  during training, already from the very first epochs. 
Thus, the probability to use unnecessary long hidden layers is quickly regularized to zero so that the \textsc{cd} algorithm only needs to update at most the parameters corresponding to the relevant $K_\text{eff}\ll K$ hidden units.   
As the probability to use a hidden layer of a certain length $z$ is conditioned on the obserbable data $\vset$, the grand-canonical theory makes sure to always assign enough hidden resources to model a given subset of the target distribution $q(\vset)$ while avoiding overfitting.
This feaure of the \vrbm is to be contrasted with the standard $\ell_p$\,--\,regularization globally implemented in the canonical Boltzmann machine at the level of the full data set.

The merits of training the suggested grand-canonical extension to the \rbm have been rigorously verified on the Ising and \textsc{mnist} data sets. In both cases, the \vrbm efficiently converges towards optimal choices for the sizes of the hidden layer leading to a very good generalization  error on previously  unseen data.
In order to get a feeling of how the \vrbm regulated itself and learned the desired features we have plotted various quantities during and after training.
In particular, we observe that the grand-canonical theory with dynamically determined chemical potential presents an extremely similar behaviour to its canonical cousin as a feature detector, once the regularization of $(K-K_\text{eff})$ hidden units has taken place in the first steps of training. 
In contrast to the trial-and-error approach mostly implemented throughout  the literature to pick some optimal size for the hidden layer of the \rbm to extract the traits from a given data set, the \vrbm managed to efficiently come to the same conclusion by dynamically regulating itself during training. 

\paragraph{Future directions.}

In this work, we have focused on a concrete ansatz Eq.~\eqref{ChemicalPotential:ConcreteAnsatz} concerning the form of the chemical potential  dictated by symmetries, intuition and some rather general assumptions on the form of the input data.
At the formal level, investigating the various phases of the grand-canonical Boltzmann system in conjunction with the extremization of the target function in the spirit of~\cite{2018JSP...172.1576D} remains an open question.
In a more applied fashion, one could relax some of our working assumptions in Section~\ref{ssc:ChemicalPotential} or try to specifically address different target data by building biases into the form of $\mu$.
Considering more versatile data sets would not only deliver stronger evidence on the self-regularizing character, but also help exhibit the flexibility of \vrbm at implementing a hidden layer of different sizes  depending on the specifics of the concrete data point being modelled. 

To clearly outline the novel aspects when training in the grand-canonical ensemble with a penalizing chemical potential, our test setting has been rather minimalistic. As a subsequent step, one could combine the theory at finite chemical potential with other  techniques of regularization (like $\ell_p$\,--\,norm) already implemented in the literature. 
Another aspect for future study is to 
extend the domain $\vset$ of the \vrbm to address multimodal distributions and data represented on the real numbers.
Last but not least, deeper networks like deep belief networks (\textsc{dbn}) and deep Boltzmann machines (\textsc{dbm}) are the next natural candidates to apply similar regularization techniques both at the level of the depth of the full network as well as the length of each hidden layer.


\subsection*{Acknowledgments}

The author would like to thank Rudolf Mayer and Deb Sarkar for very useful discussions on the topic as well as Stefan Groot Nibbelink for proof reading the manuscript. 
This work is supported within the  SBA-K1 program.
The competence center SBA Research (SBA-K1) is funded within the framework of \textsc{comet} -- Competence Centers for Excellent Technologies by \textsc{bmvit}, \textsc{bmdw}, and the federal state of Vienna, managed by the \textsc{ffg}.


\bibliographystyle{NewArXiv}
{\small
\bibliography{vRBM.bib}
}

\end{document}